\documentclass[11pt]{article}

\usepackage{authblk}
\usepackage{color}
\usepackage{ulem}
\usepackage{siunitx}
\usepackage{natbib}
\usepackage{enumitem}
\usepackage{amsmath,amssymb,amsfonts,amsthm}
\usepackage{mathtools}
\usepackage{mathabx}
\usepackage{algorithm}
\usepackage{algpseudocode}
\usepackage{setspace}
\usepackage{url}
\usepackage[a4paper,margin=20mm]{geometry}
\usepackage{caption}

\newtheorem{theorem}{Theorem}[section]
\newtheorem{lemma}[theorem]{Lemma}
\newtheorem{proposition}[theorem]{Proposition}
\newtheorem{corollary}[theorem]{Corollary}
\DeclareMathSymbol{\leq}{\mathrel}{symbols}{"14}
\DeclareMathSymbol{\geq}{\mathrel}{symbols}{"15}

\newcommand{\corraddress}[1]{\def\thecorraddress{#1}}
\newcommand{\corremail}[1]{\def\thecorremail{#1}}

\makeatletter
\newcommand{\printcorrespondence}{%
  \begingroup
  \vspace{-3em}
  \begin{center}\small
    \textbf{Correspondence:} \thecorraddress\\
    \textbf{Email:} \texttt{\thecorremail}
  \end{center}
  \endgroup
}
\g@addto@macro\maketitle{\par\vspace{-0.5em}\printcorrespondence}
\makeatother

\title{\fontsize{20pt}{24pt}\selectfont Exploring Ultra Rapid Data Assimilation Based on\\Ensemble Transform Kalman Filter\\with the Lorenz 96 Model}
\author[1]{Fumitoshi Kawasaki}
\author[2,3]{Atsushi Okazaki}
\author[2]{Kenta Kurosawa}
\author[2,3,4]{\\Shunji Kotsuki}
\affil[1]{Graduate School of Science and Engineering, Chiba University, Chiba, Japan}
\affil[2]{Center for Environmental Remote Sensing, Chiba University, Chiba, Japan}
\affil[3]{Institute for Advanced Academic Research, Chiba University, Chiba, Japan}
\affil[4]{Research Institute of Disaster Medicine, Chiba University, Chiba, Japan}
\corraddress{Fumitoshi Kawasaki, Graduate School of Science and Engineering,\\Chiba University, 1-33, Yayoicho, Inage-ku, Chiba-shi, Chiba, 263-8522, Japan.}
\corremail{fkawasaki@chiba-u.jp}
\date{}

\begin{document}
    \maketitle
    \begin{abstract}
        \noindent Ultra-rapid data assimilation (URDA) is a method that rapidly updates preemptive forecasts derived from observations without integrating a dynamical model each time additional observations become available. 
Due to its computational efficiency, we anticipate that URDA will be beneficial for application to numerical weather prediction (NWP); however, the properties of URDA in nonlinear models and its applicability to NWP have not been sufficiently elucidated.
Therefore, this study investigates the analytical properties of URDA in nonlinear models and explores inflation and localization that effectively enhance its performance, both of which are generally essential for NWP.
We first analytically demonstrate that preemptive forecasts obtained by URDA in nonlinear models are approximately equivalent, under the tangent linear approximation, to forecasts integrated from the analysis.
Furthermore, we conduct numerical experiments using the 40-variable Lorenz 96 model. 
The results show that multiplicative inflation that deliberately deflates (i.e., using an inflation factor less than $1$) the forecast ensemble perturbations used to compute the ensemble transform matrix of URDA improves forecast accuracy and inflates ensemble spread moderately. 
This is presumably attributable to the fact that deflating the forecast ensemble perturbations brings the ensemble transform matrix closer to the identity matrix and reduces the increment of the ensemble mean.
With regard to localization, we show that, although R-localization is crucial, advective localization that accounts for the advection of the influence of observations is more effective.
    \end{abstract}
    \section{INTRODUCTION}\label{sec:introduction}
The forecasting performance of numerical weather prediction (NWP) has improved significantly over the past few decades due to advances in NWP models, observational techniques, data assimilation methods, and computational technologies \citep{bauer_2015,evensen_2022,kalnay_2002}. 
However, NWP still faces challenges and limitations. One such issue is that the assimilation and forecast cycles are constrained to several hours in many global and regional models due mainly to the huge computational costs of the data assimilation process and the model forecast process \citep{etherton_2007,madaus_2015,potthast_2018,zhang_2023}.
In addition, four-dimensional data assimilation, which is employed by many operational NWP centers, requires extending the data assimilation window to incorporate many observations. This requirement may also limit high-frequency assimilation and forecast cycles. For example, in the mesoscale ensemble prediction system (MEPS) of the Japan Meteorological Agency (JMA), the frequency of 39-hour forecasts is restricted to every 6 hours. Additionally, in the medium-range ensemble forecast system (ENS) of the European Centre for Medium-Range Weather Forecasts (ECMWF), the frequency is limited to 12-hour intervals for 15-day forecasts, and to 6-hour intervals when 6-day forecasts are included.

Such infrequent assimilation and forecast cycles pose difficulties for proper decision-making in fields requiring immediacy, such as disaster prevention, aviation, and transportation. These fields require accurate prediction of rapidly evolving meteorological phenomena, including torrential rainfall associated with convection and localized squalls. However, since such meteorological phenomena occur on short time scales, significant discrepancies can arise between forecasts and actual weather conditions when assimilation and forecast cycles span several hours. If more frequent assimilation and forecast cycles were realized, the timeliness and accuracy of forecasts would be improved, thereby enhancing the reliability of action plans and preventive measures. As a result, the operational value of NWP in these fields would be further increased.

To address such issues, approaches that update forecasts rapidly without integrating dynamical models each time additional observations become available have been explored. First, \citet{etherton_2007} proposed a concept termed the preemptive forecast. The basis of this concept is to update the forecast by propagating the analysis increment obtained from data assimilation into the future and adding it to the mean of the existing forecast ensemble. That is, since this forecast update does not require integrating dynamical models, the computational cost is essentially only for the data assimilation process. Therefore, the computational costs of the preemptive forecast are much lower than those of the assimilation and forecast cycles \citep{etherton_2007}. Note that, for the preemptive forecast, it is assumed that ensemble Kalman filter \citep[EnKF;][]{evensen_1994} is used rather than variational methods such as four-dimensional variational data assimilation \citep[4DVar;][]{dimet_1986,talagrand_1987} and ensemble variational data assimilation \citep[EnVAR;][]{liu_2008}. Therefore, to be precise, the computational cost required for the preemptive forecast corresponds to that of EnKF, which is generally less expensive than the variational methods. \citet{etherton_2007} showed that the accuracy of preemptive forecasts updated based on the latest observations improved compared to the baseline 24-hour or 48-hour forecasts, using the barotropic vorticity model. Subsequently, \citet{madaus_2015} extended the preemptive forecast concept by proposing the ensemble forecast adjustment (EFA). EFA implements a preemptive forecast through the serial ensemble square root filter \citep[serial EnSRF;][]{whitaker_2002}, updating not only the forecast ensemble mean but also the forecast ensemble perturbations. \citet{madaus_2015} conducted their experiments using operational ensemble forecasts of the ECMWF and the Canadian Meteorological Centre (CMC). As a result, the preemptive forecasts of surface pressure obtained from an additional surface pressure observation showed improved accuracy up to 24 hours after the observation time in both operational centers. Furthermore, \citet{potthast_2018} proposed ultra-rapid data assimilation (URDA) based on these studies. URDA can be essentially regarded as an implementation of a preemptive forecast using the ensemble transform Kalman filter \citep[ETKF;][]{bishop_2001}. The formulation of URDA with the product of ensemble transform matrices (see Section \ref{sec:urda}) explicitly emphasizes updating preemptive forecasts sequentially with multiple observations. \citet{potthast_2018} analytically proved that the preemptive forecast with URDA is identical to the forecast from the analysis ensemble of EnKF when the system and the observation operator are linear. Furthermore, \citet{potthast_2018} demonstrated that URDA is effective even in nonlinear models by conducting experiments using the Lorenz 63 model \citep{lorenz_1963}.

Toward practical applications of URDA in NWP, this study investigates its properties in nonlinear models and proposes technical treatments to improve its performance. 
Specifically, in nonlinear models, we first demonstrate analytically that, under the tangent linear approximation, the preemptive forecast of URDA is approximately equivalent to the forecast derived from the analysis of EnKF. Although \citet{potthast_2018} numerically showed that URDA yields promising results even for nonlinear models through experiments using the Lorenz 63 model, our proof provides an analytical explanation for why URDA functions well in nonlinear models. In addition, in NWP, technical treatments such as inflation and localization are generally indispensable for data assimilation. Therefore, this study investigates the influence of inflation and localization on URDA. For this purpose, this study implements URDA using the local ensemble transform Kalman filter \citep[LETKF;][]{hunt_2007}. 
Regarding inflation, although it may seem counterintuitive, we employ an approach of deliberately deflating (i.e., applying a multiplicative inflation factor less than $1$) forecast ensemble perturbations used to compute the ensemble transform matrix of URDA.
We argue that this inflation approach is effective for improving forecast accuracy and moderately inflating the ensemble spread.
Furthermore, with respect to localization, we show that applying advective localization, which accounts for the advection of the influence of observations, instead of conventional R-localization, further improves forecast accuracy.

If URDA can be applied to NWP based on the fundamental insights gained from this investigation, more frequent forecast updates may become feasible. Apart from enabling rapid forecast updates, URDA has the following benefits, which give it high potential for applications in NWP.
\begin{itemize}
    \item Incorporating URDA within four-dimensional ensemble data assimilation systems enables high-frequency preemptive forecasts based on the latest observations while maintaining an extended data assimilation window.
    \item Unlike conventional forecasting frameworks, URDA does not require the full forecast state vector to be updated; instead, it allows selective updating of subsets of forecast states of interest \citep{madaus_2015,potthast_2018}.
    \item If the system already incorporates the LETKF scheme, URDA can be implemented easily.
\end{itemize}

In this study, we conduct numerical experiments using the 40-dimensional Lorenz 96 model \citep{lorenz_1996}, which is known as a chaotic dynamical system. The Lorenz 96 model is chosen because it can emphasize the effect of localization more easily than the three-dimensional Lorenz 63 model used in the previous studies. 

The remainder of this paper is organized as follows. In Section \ref{sec:methodology}, we describe the methodologies, including URDA, and related inflation and localization techniques. Section \ref{sec:experiment} presents the experimental design of URDA using the Lorenz 96 model. Section \ref{sec:result} shows the experimental results, and Section \ref{sec:discussion} presents the discussion. Finally, we provide a conclusion in Section \ref{sec:conclusion}.
    \section{METHODOLOGY}\label{sec:methodology}
\subsection{EnKF}
In this section, we introduce the basic formulas of EnKF. Let the dimension of the model space be $n \in \mathbb{N}$, and the dimension of the ensemble space be $m \in \mathbb{N}$. The forecast ensemble $\mathbf{X}^{f} \in \mathbb{R}^{n \times m}$, the forecast ensemble mean $\overline{\mathbf{x}}^{f} \in \mathbb{R}^{n}$, and the forecast ensemble perturbation $\delta \mathbf{X}^{f} \in \mathbb{R}^{n \times m}$ are defined using the forecast ensemble members $\mathbf{x}^{f(i)} \in \mathbb{R}^{n}$ as follows:
\begin{gather}
    \mathbf{X}^{f} \coloneqq \left[\mathbf{x}^{f(1)}, \dotsc, \mathbf{x}^{f(m)}\right],  \\
    \overline{\mathbf{x}}^{f} \coloneqq \frac{1}{m} \sum_{i=1}^{m} \mathbf{x}^{f(i)},  \\
    \delta \mathbf{X}^{f} \coloneqq \left[\mathbf{x}^{f(1)}-\overline{\mathbf{x}}^{f}, \dotsc, \mathbf{x}^{f(m)}-\overline{\mathbf{x}}^{f}\right].
\end{gather}
In addition, the error covariance matrix based on the ensemble is typically represented as follows:
\begin{equation}
    \mathbf{P}^{f} \coloneqq \frac{1}{m-1} \delta \mathbf{X}^{f}\left(\delta \mathbf{X}^{f}\right)^{\top}.  \\
\end{equation}
The analysis ensemble and analysis error covariance matrix, denoted by the superscript "$a$", are defined analogously.

The analysis ensemble mean $\overline{\mathbf{x}}^{a}$ is updated by the forecast ensemble mean $\overline{\mathbf{x}}^{f}$ as follows:
\begin{equation}
    \overline{\mathbf{x}}^{a} = \overline{\mathbf{x}}^{f}+\mathbf{K}\left(\mathbf{y}^{o}-H\left(\overline{\mathbf{x}}^{f}\right)\right)=\overline{\mathbf{x}}^{f}+\mathbf{K} \mathbf{d}^{o-f}, \label{eq:analysis_ensemble_mean}
\end{equation}
where $p \in \mathbb{N}$ denotes the dimension of the observation space, $\mathbf{y}^{o} \in \mathbb{R}^{p}$ is the observation, $H$ is the nonlinear observation operator, $\mathbf{d}^{o-f} \coloneqq \mathbf{y}^{o}-H\left(\overline{\mathbf{x}}^{f}\right) \in \mathbb{R}^{p}$ is the innovation, and $\mathbf{K} \in \mathbb{R}^{n \times p}$ is the Kalman gain.

\subsection{LETKF}
LETKF is a data assimilation method that is widely applied in many NWP studies \citep[e.g.,][]{miyoshi_2006,miyoshi_2007,szunyogh_2008} due to its advantageous properties, i.e., performing parallel computations for each grid point. For simplicity, we first describe ETKF without localization, which is the basis of LETKF. 

In ETKF, forecast error covariance matrix is expressed as follows:
\begin{equation}
    \mathbf{P}^{f}=\mathbf{Z}^{f}\left(\mathbf{Z}^{f}\right)^{\top}=\mathbf{Z}^{f} \widetilde{\mathbf{P}}^{f}\left(\mathbf{Z}^{f}\right)^{\top}, 
\end{equation}
where $\mathbf{Z}^{f} \in \mathbb{R}^{n \times m}$ is the linear map from the ensemble space to the model space, defined as follows:
\begin{equation}
    \mathbf{Z}^{f} \coloneqq \frac{1}{\sqrt{m-1}} \delta \mathbf{X}^{f}.
\end{equation}
Also, $\widetilde{\mathbf{P}}^{f} \in \mathbb{R}^{m \times m}$ is regarded as the forecast error covariance matrix in the ensemble space, which is identical to the identity matrix $\mathbf{I} \in \mathbb{R}^{m \times m}$. In the same manner, the analysis error covariance matrix $\mathbf{P}^{a}$ is expressed as follows:
\begin{equation}
    \mathbf{P}^{a}=\mathbf{Z}^{a}\left(\mathbf{Z}^{a}\right)^{\top}=\mathbf{Z}^{f} \widetilde{\mathbf{P}}^{a}\left(\mathbf{Z}^{f}\right)^{\top}, \label{eq:analysis_error_covariance_matrix}
\end{equation}
where $\mathbf{Z}^{a} \in \mathbb{R}^{n \times m}$ is defined analogously to $\mathbf{Z}^{f}$ using $\delta \mathbf{X}^{a}$.
Also, $\widetilde{\mathbf{P}}^{a} \in \mathbb{R}^{m \times m}$ is regarded as the analysis error covariance matrix in the ensemble space and is expressed as follows:
\begin{equation}
    \widetilde{\mathbf{P}}^{a}=\left[\mathbf{I}+\left(\mathbf{Y}^{f}\right)^{\top} \mathbf{R}^{-1} \mathbf{Y}^{f}\right]^{-1},
\end{equation}
where $\mathbf{R} \in \mathbb{R}^{p \times p}$ is the observation error covariance matrix, $\mathbf{Y}^{f} \in \mathbb{R}^{p \times m}$ is the linear map from the ensemble space to the observation space, defined as follows:
\begin{equation}
    \mathbf{Y}^{f}\coloneqq\mathbf{H}\mathbf{Z}^{f}\approx\frac{1}{\sqrt{m-1}}\left[H\left(\mathbf{x}^{f(1)}\right)-\overline{\mathbf{y}}^{f},\dotsc,H\left(\mathbf{x}^{f(m)}\right)-\overline{\mathbf{y}}^{f}\right],\quad\overline{\mathbf{y}}^{f}\coloneqq\frac{1}{m}\sum^{m}_{i=1}H\left(\mathbf{x}^{f(i)}\right),
\end{equation}
where $\mathbf{H} \in \mathbb{R}^{p \times n}$ is the linear map from the model space to the observation space (i.e., a linear observation operator). For example, refer to \citet{kotsuki_2020} for further details of the relationships between each space.

Here, we consider the update of the analysis ensemble mean $\overline{\mathbf{x}}^{a}$. 
The analysis increment $\mathbf{K} \mathbf{d}^{o-f}$ in Equation \eqref{eq:analysis_ensemble_mean} is expressed as follows:
\begin{align}
    \mathbf{K}\mathbf{d}^{o-f}&=\mathbf{P}^{a}\mathbf{H}^{\top}\mathbf{R}^{-1}\mathbf{d}^{o-f} \notag\\
    &\approx\mathbf{Z}^{f}\widetilde{\mathbf{P}}^{a}\left(\mathbf{Y}^{f}\right)^{\top}\mathbf{R}^{-1}\mathbf{d}^{o-f} \notag\\
    &=\mathbf{Z}^{f}\mathbf{w}, 
    \quad \mathbf{w} \coloneqq \widetilde{\mathbf{P}}^{a}\left(\mathbf{Y}^{f}\right)^{\top}\mathbf{R}^{-1}\mathbf{d}^{o-f}.
    \label{eq:analysis_increment}
\end{align}
Therefore, from Equations \eqref{eq:analysis_ensemble_mean}, and \eqref{eq:analysis_increment}, the analysis ensemble mean $\overline{\mathbf{x}}^{a}$ can be expressed as follows:
\begin{equation}
    \overline{\mathbf{x}}^{a}=\overline{\mathbf{x}}^{f}+\mathbf{Z}^{f} \mathbf{w}. \label{eq:ensemble_update_of_mean}
\end{equation}
Subsequently, we consider the update of the analysis ensemble perturbation $\delta \mathbf{X}^{a}$. Since an error covariance matrix is a positive semidefinite matrix, from Equation \eqref{eq:analysis_error_covariance_matrix}, we have
\begin{equation}
    \mathbf{P}^{a}=\left[\mathbf{Z}^{f}\left(\widetilde{\mathbf{P}}^{a}\right)^{\frac{1}{2}} \mathbf{U} \right]\left[\mathbf{Z}^{f}\left(\widetilde{\mathbf{P}}^{a}\right)^{\frac{1}{2}} \mathbf{U} \right]^{\top},
\end{equation}
where $\mathbf{U}$ is a certain orthogonal matrix.
Here, as one reasonable choice \citep{duc_2020}, we select $\mathbf{U} = \mathbf{I}$, which gives
\begin{equation}
    \mathbf{Z}^{a}=\mathbf{Z}^{f}\left(\widetilde{\mathbf{P}}^{a}\right)^{\frac{1}{2}}.
\end{equation}
The analysis ensemble perturbation $\delta \mathbf{X}^{a}$ is then expressed as a linear combination of the forecast ensemble perturbation $\delta \mathbf{X}^{f}$:
\begin{equation}
    \delta \mathbf{X}^{a}=\delta \mathbf{X}^{f}\left(\widetilde{\mathbf{P}}^{a}\right)^{\frac{1}{2}}=\delta \mathbf{X}^{f} \mathbf{W}, \quad \mathbf{W} \coloneqq \left(\widetilde{\mathbf{P}}^{a}\right)^{\frac{1}{2}}. \label{eq:ensemble_update_of_perturbation}
\end{equation}
Therefore, from Equations \eqref{eq:ensemble_update_of_mean} and \eqref{eq:ensemble_update_of_perturbation}, the update equation of the analysis ensemble $\mathbf{X}^{a}$ is expressed by:
\begin{align}
    \mathbf{X}^{a}&=\overline{\mathbf{x}}^{a} \mathbf{1}^{\top}+\delta \mathbf{X}^{a} \notag \\
    & =\overline{\mathbf{x}}^{f} \mathbf{1}^{\top}+\mathbf{Z}^{f}\left(\mathbf{w} \mathbf{1}^{\top}+\sqrt{m-1} \mathbf{W}\right) \notag \\
    & =\overline{\mathbf{x}}^{f} \mathbf{1}^{\top}+\mathbf{Z}^{f} \widehat{\mathbf{W}}, \quad \mathbf{1} \coloneqq \left[1, \dotsc, 1\right]^{\top}, \quad \widehat{\mathbf{W}} \coloneqq \mathbf{w} \mathbf{1}^{\top}+\sqrt{m-1} \mathbf{W}. \label{eq:ensemble_update_of_ensemble}
\end{align}
Furthermore, the analysis ensemble $\mathbf{X}^{a}$ can be expressed in terms of the properties of $\mathbf{w}$ and $\mathbf{W}$ \citep[e.g.,][]{potthast_2018} as follows:
\begin{equation}
    \mathbf{X}^{a}=\frac{1}{\sqrt{m-1}} \mathbf{X}^{f} \widehat{\mathbf{W}}=\mathbf{X}^{f} \widecheck{\mathbf{W}}, \quad \widecheck{\mathbf{W}} \coloneqq \frac{1}{\sqrt{m-1}} \widehat{\mathbf{W}}.
\end{equation}

Now, we consider the update equations of LETKF, which incorporates localization. In LETKF, the analysis ensemble is computed for each grid point using a subset of observations selected based on a localization scheme, as follows:
\begin{align}
    \mathbf{x}_g^{a}
    &= \mathbf{x}_g^{f}\,\widecheck{\mathbf{W}}^{loc} \notag \\
    &= \mathbf{x}_g^{f}\Biggl[
        \frac{1}{\sqrt{m-1}}\bigl(\mathbf{Y}^{f,loc}\bigr)^{\top}\Bigl[
        \mathbf{Y}^{f,loc}\bigl(\mathbf{Y}^{f,loc}\bigr)^{\top}
        + \mathbf{R}^{loc}\Bigr]^{-1}
        \mathbf{d}^{o-f,loc}\mathbf{1}^{\top} \notag \\
    &\hspace{12em}
        +\Bigl[\mathbf{I}
        +\bigl(\mathbf{Y}^{f,loc}\bigr)^{\top}
        \bigl(\mathbf{R}^{loc}\bigr)^{-1}
        \mathbf{Y}^{f,loc}
        \Bigr]^{-1/2}
        \Biggr],
\end{align}
where the subscript "$g$" denotes the $g$-th grid point (i.e., $\mathbf{x}_{g}^{a}, \mathbf{x}_{g}^{f} \in \mathbb{R}^{m}$ are row vectors), and the superscript "$loc$" denotes that the vector or matrix is based on a subset of the observations selected by the localization scheme at the $g$-th grid point. For example, refer to \citet{hunt_2007} for details of LETKF.

\subsection{URDA}\label{sec:urda}
\begin{figure}[H]
    \centering
    \includegraphics[width=13cm]{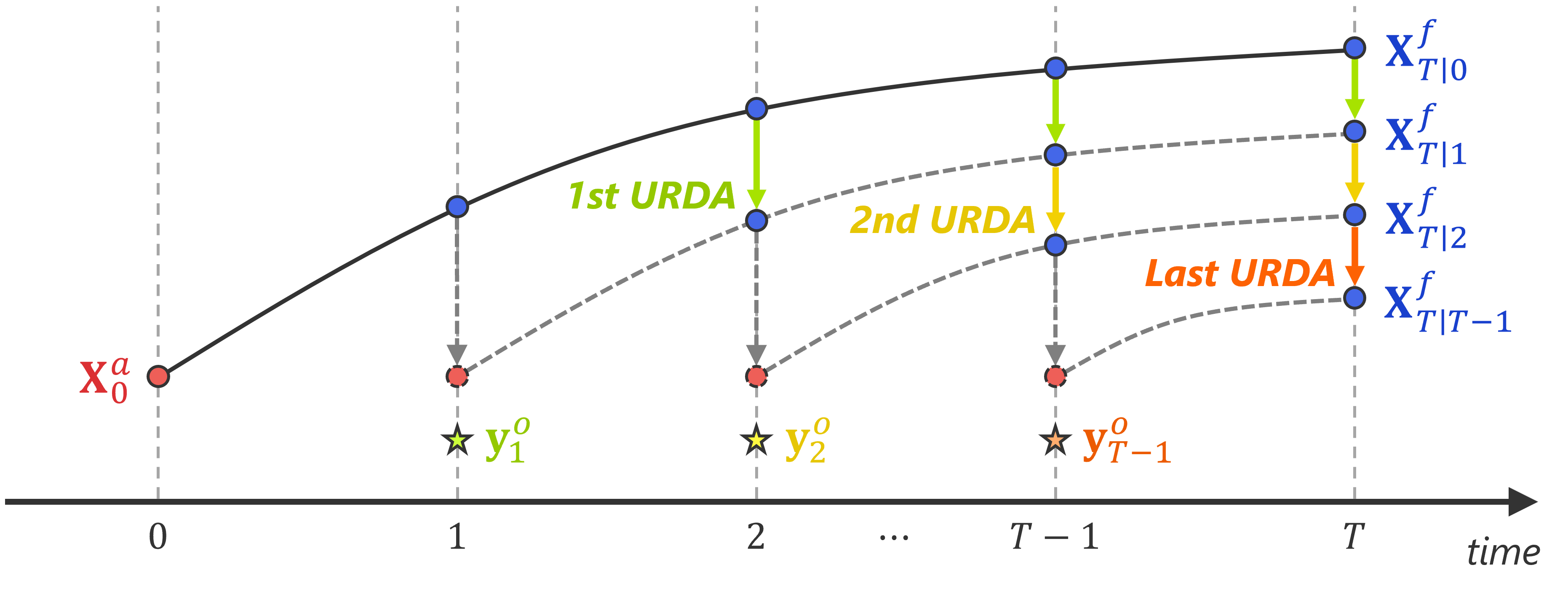}
    \caption{The concept of URDA. A situation is assumed in which the baseline forecast already exists from time $0$ to time $T$, and additional observations become available at each subsequent time step. URDA sequentially updates the preemptive forecasts using the existing forecast ensembles and the additional observations.}
    \label{fig:concept_urda}
\end{figure}

We first clarify the problem setting for URDA. As shown in Figure \ref{fig:concept_urda}, we assume a situation in which an ensemble forecast with a dynamical model (hereafter called the baseline forecast) has already been performed from time $0$ to time $T$, and then additional observations become available at each subsequent time step. For the sake of simplicity, we first consider URDA without localization.

As shown in Figure \ref{fig:concept_urda}, URDA sequentially updates preemptive forecasts based on existing forecast ensembles and additional available observations (hereafter referred to as URDA forecasts). That is, unlike ordinary data assimilation, URDA does not require computing an analysis, but only updating the forecasts. In particular, when the model and the observation operator are linear, \citet{potthast_2018} demonstrated that the following equation holds:
\begin{equation}
    \mathbf{X}_{k \mid j}^{f}=\mathbf{X}_{k \mid 0}^{f} \widecheck{\mathbf{W}}_{1} \dotsm \widecheck{\mathbf{W}}_{j}=\mathbf{X}_{k \mid 0}^{f} \widecheck{\mathbf{W}}_{j}^{prod}, \quad \widecheck{\mathbf{W}}_{j}^{prod} \coloneqq \widecheck{\mathbf{W}}_{1} \dotsm \widecheck{\mathbf{W}}_{j}, \label{eq:urda_linear}
\end{equation}
where subscript $k \mid j \ \left(0 \leq j<T, j<k \leq T\right)$ denotes that the forecast is performed from time $j$ to time $k$. In this study, we refer to $j$ as the forecast reference time and $k$ as the forecast lead time. Equation \eqref{eq:urda_linear} indicates that the forecast ensemble $\mathbf{X}_{k \mid j}^{f}$ can be updated sequentially by applying the ensemble transform matrix $\widecheck{\mathbf{W}}_{j}$ from the right, when an observation $\mathbf{y}_{j}^{o}$ is available at forecast reference time $j$. However, for nonlinear models, the effectiveness of URDA has been demonstrated only through numerical experiments by \citet{potthast_2018}.

Therefore, this study analytically demonstrates that Equation \eqref{eq:urda_linear} holds approximately even for nonlinear models. First, we provide the following lemma to apply linear approximation.

\begin{lemma}\label{lem:analysis_ensemble_member}
    At time $j$, the analysis ensemble member $\mathbf{x}_{j}^{a(i)}$ is represented as follows in terms of the forecast ensemble member $\mathbf{x}_{j \mid j-1}^{f(i)}$, the forecast ensemble perturbation $\delta \mathbf{X}_{j \mid j-1}^{f}$, and the ensemble transform matrix $\widecheck{\mathbf{W}}_{j}$ :
    \begin{equation}
        \mathbf{x}_{j}^{a(i)}=\mathbf{x}_{j \mid j-1}^{f(i)}+\sum_{l=1}^{m} \delta \mathbf{x}_{j \mid j-1}^{f(l)} \omega_{j, l i}, \label{eq:lemma1}
    \end{equation}
    where $\omega_{j, l i}$ is the $l$-th row and $i$-th column element of $\boldsymbol{\Omega}_{j} \coloneqq \widecheck{\mathbf{W}}_{j}-\mathbf{I}$.
\end{lemma}

\begin{proof}
    From Equation \eqref{eq:ensemble_update_of_ensemble}, the following equation holds:
    \begin{align}
        \mathbf{X}_{j}^{a} & =\overline{\mathbf{x}}_{j \mid j-1}^{f} \mathbf{1}^{\top}+\mathbf{Z}_{j \mid j-1}^{f} \widehat{\mathbf{W}}_{j} \notag \\
        & =\overline{\mathbf{x}}_{j \mid j-1}^{f} \mathbf{1}^{\top}+\delta \mathbf{X}_{j \mid j-1}^{f} \widecheck{\mathbf{W}}_{j}+\delta \mathbf{X}_{j \mid j-1}^{f}-\delta \mathbf{X}_{j \mid j-1}^{f} \notag \\
        & =\mathbf{X}_{j \mid j-1}^{f}+\delta \mathbf{X}_{j \mid j-1}^{f} \boldsymbol{\Omega}_{j}. 
    \end{align}
    Considering the $i$-th column of $\mathbf{X}_{j}^{a}$, we obtain Equation \eqref{eq:lemma1}.
\end{proof}
Then, based on Lemma \ref{lem:analysis_ensemble_member}, we provide an approximate equation of the updated forecast ensemble $\mathbf{X}_{k \mid j}^{f}$ derived from the baseline forecast $\mathbf{X}_{k \mid 0}^{f}$.

\begin{lemma}\label{lem:forecast_ensemble}
    The forecast ensemble $\mathbf{X}_{k \mid j}^{f}$ for forecast lead time $k$ updated at forecast reference time $j$ is represented as follows by the baseline forecast $\mathbf{X}_{k \mid 0}^{f}$, the forecast ensemble perturbations $\left\{\delta \mathbf{X}_{k \mid h-1}^{f}\right\}_{h=1}^{j}$ and the matrices $\left\{\boldsymbol{\Omega}_{h}\right\}_{h=1}^{j}$:
    \begin{equation}
        \mathbf{X}_{k \mid j}^{f} \approx \mathbf{X}_{k \mid 0}^{f}+\sum_{h=1}^{j} \delta \mathbf{X}_{k \mid h-1}^{f} \boldsymbol{\Omega}_{h}. \label{eq:lemma2}
    \end{equation}
\end{lemma}

\begin{proof}
    From Lemma \ref{lem:analysis_ensemble_member}, the following equation holds:
    \begin{alignat}{2}
        &\mathbf{x}_{k \mid 1}^{f(i)} && = M_{k \mid 1}\left(\mathbf{x}_{1}^{a(i)}\right) \notag \\
        &&& = M_{k \mid 1}\left(\mathbf{x}_{1 \mid 0}^{f(i)}+\sum_{l=1}^{m} \delta \mathbf{x}_{1 \mid 0}^{f(l)} \omega_{1, l i}\right) \notag \\
        &&& \approx M_{k \mid 1}\left(\mathbf{x}_{1 \mid 0}^{f(i)}\right)+\sum_{l=1}^{m} \mathbf{M}_{k \mid 1} \delta \mathbf{x}_{1 \mid 0}^{f(l)} \omega_{1, l i} \notag \\
        &&& = \mathbf{x}_{k \mid 0}^{f(i)}+\sum_{l=1}^{m} \delta \mathbf{x}_{k \mid 0}^{f(l)} \omega_{1, l i}.
    \end{alignat}
    Considering the above equation for all $i=1,\ldots,m$, the forecast ensemble $\mathbf{X}_{k \mid 1}^{f}$ is represented by:
    \begin{equation}
        \mathbf{X}_{k \mid 1}^{f} \approx \mathbf{X}_{k \mid 0}^{f}+\delta \mathbf{X}_{k \mid 0}^{f} \boldsymbol{\Omega}_{1}.
    \end{equation}
    Therefore, Equation \eqref{eq:lemma2} holds for $j=1$. We assume that Equation \eqref{eq:lemma2} holds for $j=t$. Then, considering the case $j=t+1$, the following equation holds:
    \begin{alignat}{2}
        &\mathbf{x}_{k \mid t+1}^{f(i)} && =M_{k \mid t+1}\left(\mathbf{x}_{t+1}^{a(i)}\right) \notag \\
        &&& =M_{k \mid t+1}\left(\mathbf{x}_{t+1 \mid t}^{f(i)}+\sum_{l=1}^{m} \delta \mathbf{x}_{t+1 \mid t}^{f(l)} \omega_{t+1, l i}\right) \notag \\
        &&& \approx M_{k \mid t+1}\left(\mathbf{x}_{t+1 \mid t}^{f(i)}\right)+\sum_{l=1}^{m} \mathbf{M}_{k \mid t+1} \delta \mathbf{x}_{t+1 \mid t}^{f(l)} \omega_{t+1, l i} \notag \\
        &&& =\mathbf{x}_{k \mid t}^{f(i)}+\sum_{l=1}^{m} \delta \mathbf{x}_{k \mid t}^{f(l)} \omega_{t+1, l i} \notag \\
        &&& \approx \mathbf{x}_{k \mid 0}^{f(i)}+\sum_{h=1}^{t} \sum_{l=1}^{m} \delta \mathbf{x}_{k \mid h-1}^{f(l)} \omega_{h, l i}+\sum_{l=1}^{m} \delta \mathbf{x}_{k \mid t}^{f(l)} \omega_{t+1, l i} \notag \\
        &&& =\mathbf{x}_{k \mid 0}^{f(i)}+\sum_{h=1}^{t+1} \sum_{l=1}^{m} \delta \mathbf{x}_{k \mid h-1}^{f(l)} \omega_{h, l i}.
    \end{alignat}
    Applying the above equation for all $i=1,\ldots,m$, the forecast ensemble $\mathbf{X}_{k \mid t+1}^{f}$ is given by:
    \begin{equation}
        \mathbf{X}_{k \mid t+1}^{f} \approx \mathbf{X}_{k \mid 0}^{f}+\sum_{h=1}^{t+1} \delta \mathbf{X}_{k \mid h-1}^{f} \boldsymbol{\Omega}_{h}.
    \end{equation}
    Therefore, because it also holds for $j=t+1$, Equation \eqref{eq:lemma2} holds for all $j$.
\end{proof}
Here, the approximation in Equation \eqref{eq:lemma2} denotes tangent linear approximations around $\mathbf{x}^{f(i)}_{h\mid h-1}$ for each $h=1,\ldots,j$ and $i=1,\ldots,m$. Therefore, it is suggested that errors in the tangent linear approximation accumulate as $j$ increases.

Subsequently, to represent the updated forecast ensemble $\mathbf{X}_{k \mid j}^{f}$ by the product of ensemble transform matrices, we consider the following lemma.

\begin{lemma}\label{lem:sum2prod}
    For the baseline forecast $\mathbf{X}_{k \mid 0}^{f}$, the forecast ensemble perturbations $\left\{ \delta\mathbf{X}^{f}_{k\mid h-1}\right\} ^{j}_{h=1}$, the matrices $\left\{\boldsymbol{\Omega}_{h}\right\} ^{j}_{h=1}$, and the ensemble transform matrices $\left\{\widecheck{\mathbf{W}}_{h}\right\} ^{j}_{h=1}$, the following equation holds:
    \begin{equation}
        \mathbf{X}_{k \mid 0}^{f}+\sum_{h=1}^{j} \delta \mathbf{X}_{k \mid h-1}^{f} \boldsymbol{\Omega}_{h} \approx \mathbf{X}_{k \mid 0}^{f} \widecheck{\mathbf{W}}_{1} \dotsm \widecheck{\mathbf{W}}_{j}. \label{eq:lemma3}
    \end{equation}
\end{lemma}

\begin{proof}
    Noting that $\mathbf{1}^{\top} \widecheck{\mathbf{W}}_{1}=\mathbf{1}^{\top}$, the following equation is derived:
    \begin{align}
        \mathbf{X}_{k \mid 0}^{f}+\delta \mathbf{X}_{k \mid 0}^{f} \boldsymbol{\Omega}_{1} & =\mathbf{X}_{k \mid 0}^{f}+\delta \mathbf{X}_{k \mid 0}^{f}\left(\widecheck{\mathbf{W}}_{1}-\mathbf{I}\right) \notag \\
        & =\overline{\mathbf{x}}_{k \mid 0}^{f} \mathbf{1}^{\top} \widecheck{\mathbf{W}}_{1}+\delta \mathbf{X}_{k \mid 0}^{f} \widecheck{\mathbf{W}}_{1} \notag \\
        & =\mathbf{X}_{k \mid 0}^{f} \widecheck{\mathbf{W}}_{1}.
    \end{align}
    Therefore, Equation \eqref{eq:lemma3} holds for $j=1$. We assume that Equation \eqref{eq:lemma3} holds for $j=t$. Then, considering the case $j=t+1$, the following equation holds from Lemma \ref{lem:forecast_ensemble}:
    \begin{align}
        \mathbf{X}_{k \mid 0}^{f}+\sum_{h=1}^{t+1} \delta \mathbf{X}_{k \mid h-1}^{f} \boldsymbol{\Omega}_{h} & =\mathbf{X}_{k \mid 0}^{f}+\sum_{h=1}^{t} \delta \mathbf{X}_{k \mid h-1}^{f} \boldsymbol{\Omega}_{h}+\delta \mathbf{X}_{k \mid t}^{f} \boldsymbol{\Omega}_{t+1} \notag \\
        & \approx \mathbf{X}_{k \mid t}^{f}+\delta \mathbf{X}_{k \mid t}^{f}\left(\widecheck{\mathbf{W}}_{t+1}-\mathbf{I}\right) \notag \\
        & =\overline{\mathbf{x}}_{k \mid t}^{f} \mathbf{1}^{\top} \widecheck{\mathbf{W}}_{t+1}+\delta \mathbf{X}_{k \mid t}^{f} \widecheck{\mathbf{W}}_{t+1} \notag \\
        & =\mathbf{X}_{k \mid t}^{f} \widecheck{\mathbf{W}}_{t+1} \notag \\
        & \approx\left[\mathbf{X}_{k \mid 0}^{f}+\sum_{h=1}^{t} \delta \mathbf{X}_{k \mid h-1}^{f} \boldsymbol{\Omega}_{h}\right] \widecheck{\mathbf{W}}_{t+1} \notag \\
        & \approx \mathbf{X}_{k \mid 0}^{f} \widecheck{\mathbf{W}}_{1} \dotsm \widecheck{\mathbf{W}}_{t+1}.
    \end{align}
    Therefore, because it also holds for $j=t+1$, Equation \eqref{eq:lemma3} holds for all $j$.
\end{proof}
The approximation in Equation \eqref{eq:lemma3} originates from Equation \eqref{eq:lemma2} and may involve tangent linear approximations around $\mathbf{x}^{f(i)}_{h\mid h-1}$ for each $h=1,\ldots,j$ and $i=1,\ldots,m$.
However, since this approximation is employed solely as a bidirectional transformation between $\mathbf{X}^{f}_{k\mid0}+\sum^{t}_{h=1}\delta\mathbf{X}^{f}_{k\mid h-1}\boldsymbol{\Omega}_{h}$ and $\mathbf{X}^{f}_{k\mid t}$, the degree of approximation may be negligibly small in practice despite its mathematical presence. 
Indeed, a numerical comparison of both sides of Equation \eqref{eq:lemma3} for an arbitrary case under the standard URDA settings of this study using the Lorenz 96 model confirmed agreement to $12$ decimal places.

These lemmas lead to the following theorem.

\begin{theorem}\label{th:urda_nonlinear}
    In a nonlinear model, the forecast ensemble $\mathbf{X}_{k \mid j}^{f}$ for forecast lead time $k$ at forecast reference time $j$ is expressed as follows using the baseline forecast $\mathbf{X}_{k \mid 0}^{f}$ and the ensemble transform matrices $\left\{\widecheck{\mathbf{W}}_{h}\right\} ^{j}_{h=1}$ :
    \begin{equation}
        \mathbf{X}_{k \mid j}^{f} \approx \mathbf{X}_{k \mid 0}^{f} \widecheck{\mathbf{W}}_{1} \dotsm \widecheck{\mathbf{W}}_{j}=\mathbf{X}_{k \mid 0}^{f} \widecheck{\mathbf{W}}_{j}^{prod}. \label{eq:theorem1} 
    \end{equation}
\end{theorem}

\begin{proof}
    This follows immediately from Lemmas \ref{lem:forecast_ensemble} and \ref{lem:sum2prod}.
\end{proof}
Note that $\widecheck{\mathbf{W}}_{j}^{prod}$ can be decomposed into $\mathbf{w}_{j}^{prod}$ and $\mathbf{W}_{j}^{prod}$ as follows:
\begin{equation}
    \widecheck{\mathbf{W}}_{j}^{prod}=\frac{1}{\sqrt{m-1}} \mathbf{w}_{j}^{prod} \mathbf{1}^{\top}+\mathbf{W}_{j}^{prod}, \label{eq:decompose_of_ensemble_transform_matrix}
\end{equation}
where $\mathbf{w}_{j}^{prod} \coloneqq \sum_{t=1}^{j} \mathbf{W}_{t-1}^{prod} \mathbf{w}_{t}$ and $\mathbf{W}_{0}^{prod} \coloneqq \mathbf{I}$.
Also, $\mathbf{w}_{j}^{prod}$ and $\mathbf{W}_{j}^{prod}$ can be calculated using the ensemble transform matrix $\widecheck{\mathbf{W}}_{j}^{prod}$ as follows:
\begin{gather}
    \mathbf{w}_{j}^{prod}=\frac{\sqrt{m-1}}{m}\left(\widecheck{\mathbf{W}}_{j}^{prod}-\mathbf{I}\right) \mathbf{1}, \label{eq:ensemble_analysis_increment_prod} \\
    \mathbf{W}_{j}^{prod}=\widecheck{\mathbf{W}}_{j}^{prod}-\frac{1}{m}\left(\widecheck{\mathbf{W}}_{j}^{prod}-\mathbf{I}\right) \mathbf{J}, \quad \mathbf{J} \coloneqq \mathbf{1}\mathbf{1}^{\top}. \label{eq:ensemble_transform_matrix_prod}
\end{gather}
For the derivation of Equations \eqref{eq:decompose_of_ensemble_transform_matrix} -- \eqref{eq:ensemble_transform_matrix_prod}, refer to Appendix \ref{sec:appendix_a}. 

Introducing localization into URDA implemented by LETKF, Equation \eqref{eq:theorem1} is expressed as follows: 
\begin{equation}
    \mathbf{x}_{k \mid j, g}^{f} \approx \mathbf{x}_{k \mid 0, g}^{f} \widecheck{\mathbf{W}}_{1}^{loc} \dotsm \widecheck{\mathbf{W}}_{j}^{loc}=\mathbf{x}_{k \mid 0, g}^{f} \widecheck{\mathbf{W}}_{j}^{prod,loc}. \label{eq:urda_nonlinear_localization}
\end{equation}

\subsection{Technical treatments for URDA}
\subsubsection{Inflation}
In conventional data assimilation studies, multiplicative inflation \citep{anderson_1999}, relaxation to prior perturbations \citep[RTPP;][]{zhang_2004}, and relaxation to prior spread \citep[RTPS;][]{whitaker_2012} have been widely used as basic inflation methods.
However, whereas the ensemble transform matrix of URDA possesses the property $\mathbf{1}^{\top} \widecheck{\mathbf{W}}_{j}^{prod}=\mathbf{1}^{\top}$ (see Appendix \ref{sec:appendix_a}), RTPS violates this property because it is regarded as multiplying the ensemble transform matrix by a constant factor. 
Consequently, RTPS may adversely affect URDA.
Indeed, in our preliminary experiments, RTPS induced numerical divergence in many cases. 
Therefore, this study employs multiplicative inflation and RTPP as candidate inflation methods for URDA.

In the multiplicative inflation in URDA, several options are possible for inflating the forecast ensemble perturbations.
From Equations \eqref{eq:theorem1} and \eqref{eq:decompose_of_ensemble_transform_matrix}, the forecast perturbation of URDA can be expressed as $\delta\mathbf{X}^{f}_{k\mid j} = \delta\mathbf{X}^{f}_{k\mid 0} \mathbf{W}^{prod}_{j}$. 
Therefore, one possible approach is to inflate as $\delta\mathbf{X}^{f,multi}_{k\mid j}=\alpha^{multi}\cdot\delta\mathbf{X}^{f}_{k\mid0}\mathbf{W}^{prod}_{j}$ where $\alpha^{multi}$ is the multiplicative inflation factor and the superscript $multi$ denotes that multiplicative inflation is applied.
However, this approach was found to be ineffective in our preliminary experiments.
Instead, we note that the ensemble transform matrix $\widecheck{\mathbf{W}}_{j}$ is computed using $\mathbf{Y}^{f}_{j\mid j-1}$. 
In this study, multiplicative inflation is therefore applied to the ensemble transform matrix by replacing $\mathbf{Y}^{f}_{j\mid j-1}$ with $\alpha^{multi} \cdot \mathbf{Y}^{f}_{j\mid j-1}$ as follows:
\begin{gather}
    \widecheck{\mathbf{W}}^{multi}_{j}=\frac{1}{\sqrt{m-1}}\mathbf{w}^{multi}_{j}\mathbf{1}^{\top}+\mathbf{W}^{multi}_{j},\\
    \mathbf{w}^{multi}_{j}=\alpha^{multi}\cdot\widetilde{\mathbf{P}}^{a,multi}_{j}\left(\mathbf{Y}^{f}_{j\mid j-1}\right)^{\top}\mathbf{R}^{-1}\mathbf{d}^{o-f}_{j},\label{eq:ensemble_analysis_increment_inf}\\
    \mathbf{W}^{multi}_{j}=\left(\widetilde{\mathbf{P}}^{a,multi}_{j}\right)^{\frac{1}{2}}, \label{eq:ensemble_transform_matrix_inf}
\end{gather}
where $\widetilde{\mathbf{P}}^{a,multi}_{j}$ is represented as follows:
\begin{equation}
    \widetilde{\mathbf{P}}^{a,multi}_{j}=\left[\mathbf{I}+\left(\alpha^{multi}\right)^{2}\cdot\left(\mathbf{Y}^{f}_{j\mid j-1}\right)^{\top}\mathbf{R}^{-1}\mathbf{Y}^{f}_{j\mid j-1}\right]^{-1}. \label{eq:ensemble_analysis_error_covariance_matrix_inf}
\end{equation}
That is, in this approach, note that multiplicative inflation is applied only to the ensemble transform matrix $\widecheck{\mathbf{W}}_{j}$.

RTPP is applied to the ensemble transform matrix $\mathbf{W}_{j}$ as follows:
\begin{equation}
    \mathbf{W}^{rtpp}_{j}=\left(1-\alpha^{rtpp}\right)\mathbf{W}_{j}+\alpha^{rtpp}\mathbf{I}, \label{eq:rtpp}
\end{equation}
where $\alpha^{rtpp}$ is the inflation factor of RTPP and the superscript $rtpp$ denotes that RTPP is applied.

\subsubsection{Localization}
The influence of observations is advected as time evolves.
To account for such advection of the influence of observations when applying R-localization, the localization parameters such as the localization center should be varied adaptively with time.

Although it is not obvious how the localization should be varied adaptively over time, in this study we employ the simple advective localization proposed by \citet{kalnay_2012} in their study of ensemble forecast sensitivity to observation (EFSO).
In this advective localization, the localization center, which coincides with the analysis grid point in standard R-localization, is shifted at a constant velocity with time (see Figure 5b of \citet{kalnay_2012}). 
That is, in URDA, the localization center is shifted based on $k-j$, the time elapsed from the forecast reference time $j$ to the forecast lead time $k$.

Ideally, it would be desirable to compute the ensemble transform matrix $\widecheck{\mathbf{W}}_{j}$ by shifting the localization center for each forecast lead time $k$. 
However, this computation requires a significant computational cost.
We therefore divide the range of possible forecast lead times $(0, T]$ into $l$ time slots $s_{i}=\left[t_{i},T_{i}\right]$ ($i=1,\ldots,l$), and compute the ensemble transform matrix $\widecheck{\mathbf{W}}_{j}$ once per slot. The time slots are mutually disjoint and collectively cover the entire range. 
Specifically, when a forecast lead time $k$ satisfies $k \in s_{i}$, the localization center is shifted based on $T_i - j$, and the ensemble transform matrix $\widecheck{\mathbf{W}}_{j}$ is computed once for slot $s_{i}$.
This ensemble transform matrix $\widecheck{\mathbf{W}}_{j}$ is then reused for all subsequent forecast lead times $k$ satisfying $k \in s_{i}$.
Reducing the number of time slots $l$ is expected to lower the computational cost but it may simultaneously introduce a departure from the desirably shifted localization center, or lead to discontinuities in the forecast ensemble across forecast lead times $k$.

    \section{EXPERIMENT}\label{sec:experiment}
\subsection{The Lorenz 96 model}
In this study, numerical experiments with URDA are conducted using the 40-variable Lorenz 96 model, a chaotic dynamical system. The Lorenz 96 model can be regarded as mimicking certain physical quantities distributed uniformly on the same latitude circle \citep{lorenz_1998a}. The equations of the Lorenz 96 model are represented as differential equations composed of advection terms, dissipation terms, and forcing terms that are intrinsic to meteorological phenomena, as follows:
\begin{equation}
    \frac{\mathrm{d} x_{g}}{\mathrm{d} t}=\left(x_{g+1}-x_{g-2}\right) x_{g-1}-x_{g}+F, \quad g=1, \dotsc, n,
\end{equation}
where the dimension of the model space is $n=40$ and the forcing term is set to $F=8.0$ to induce chaotic behavior in the system. The system is numerically integrated using a fourth-order Runge--Kutta scheme. One time step in the integration is defined as a dimensionless time unit, $\mathrm{d} t=0.01$. In the Lorenz 96 model with $F=8.0$, one unit of time corresponds to 5 days (i.e., $\mathrm{d} t=0.01$ unit of time corresponds to 1.2 hours) based on the discussion of error doubling time by \citet{lorenz_1998a}.

\subsection{Experimental design of URDA}
The experimental design of URDA is outlined as follows.
\begin{enumerate}
    \item An observing system simulation experiment (OSSE) is performed to save the true states for evaluation of URDA and the analysis ensembles as the initial values of URDA.
    \item URDA experiments are conducted using observations and baseline forecasts generated from the saved true states and analysis ensembles, respectively.
\end{enumerate}

As preparation for the URDA experiments, the OSSE procedure is described. First, a $1$-year spin-up is performed, followed by a $2$-year plus $30$-day integration of the Lorenz 96 model, which is saved as the true state. Then, observations for the OSSE are generated by adding Gaussian noise $\mathcal{N}(0,1)$ to the true state. Here, the dimension of the observation space is $p=n$, the observation interval is $\Delta t=0.05$, and the observation operator is the identity matrix. Subsequently, the assimilation and forecast cycles are performed. LETKF is employed as the data assimilation method, with an ensemble size of $10$ members. The multiplicative inflation factor $\alpha^{multi}$ and the localization scale $\sigma$ were tuned, with values of $\alpha^{multi}=1.03$ and $\sigma=5.5$ selected, respectively. The forecast ensemble at the initial time was generated by adding Gaussian noise $\mathcal{N}(0,1)$ to the true state, similar to the observations. To avoid using poor-quality analysis ensembles, the data from the first 30 days is removed. Accordingly, the true state from the first 30 days is also removed. Furthermore, to provide independent initial values for the URDA experiments, 293 steps of true states and analysis ensembles extracted every 60 hours from the remaining $2$ years are saved.

Then, the procedure for the URDA experiments is explained. From the 293 steps of the true states and analysis ensembles saved in the OSSE, the $t$-th step is extracted and used as the true state $\mathbf{x}_{0}^{tru}$ and the analysis ensemble $\mathbf{X}_{0}^{a}$ at the initial time in the time interval $\left[0,\ T\right]$. 
The end of the time interval $T$ is set to $7$ days, within which the tangent linear approximation is expected to be functional.
Subsequently, the baseline forecast $\mathbf{X}_{k \mid 0}^{f} \left(0<k \leq T\right)$ is computed by integrating the Lorenz 96 model from the analysis ensemble $\mathbf{X}_{0}^{a}$. In the same manner, the true state $\mathbf{x}_{j}^{tru} \left(0<j \leq T\right)$ is computed. Then, the observation $\mathbf{y}_{j}^{o}$ used in the URDA experiments is generated by adding Gaussian noise $\mathcal{N}(0,1)$ to the true state $\mathrm{x}_{j}^{tru}$. 
As URDA assumes the assimilation of high-frequency observations, the observation interval is set to $\Delta t=0.01$. Other settings related to observation, such as the observation dimension and observation operator, are basically consistent with those used in the previous OSSE.
Then, URDA is performed using the baseline forecast $\mathbf{X}_{k \mid 0}^{f}$ and the observation $\mathbf{y}_{j}^{o}$. 
This process is applied to all stored data for $t=1, \dotsc, 293$ steps, and statistical evaluations are eventually conducted for performance metrics described later.

\subsection{Technical treatments}
Typically, the multiplicative inflation factor $\alpha^{multi}$ is a real number greater than or equal to $1$; however, in the URDA experiments, it is allowed to take values less than $1$.
This is because, as will be discussed in detail in Section \ref{sec:discussion}, our experiments showed that values of $\alpha^{multi}\leq1$ yielded better results.
That is, although referred to here as multiplicative inflation for convenience, this term also encompasses deflation in this study.
This can also be interpreted as inflating the observation error covariance matrix $\mathbf{R}$. 
In practice, inflating $\mathbf{R}$ yielded results similar to those obtained by deflating $\mathbf{Y}^{f}_{j\mid j-1}$ in our preliminary experiment.
However, when we adjusted these to have the same ensemble perturbation under a constant inflation factor and compared them, the forecast accuracy was not identical; the deflation of $\mathbf{Y}^{f}_{j\mid j-1}$ was shown to be superior in various respects.
These differences arise from the asymmetry of the inflation factors applied to $\mathbf{w}_j$ and $\mathbf{W}_{j}$. Specifically, $\mathbf{w}_j$ obtained by inflating $\mathbf{R}$ is scaled by a factor of $\alpha^{multi}$ relative to that obtained by deflating $\mathbf{Y}^{f}_{j\mid j-1}$.
When $\alpha^{multi}=0$, the URDA forecast is identical to the baseline forecast.
Therefore, $\alpha^{multi}$ is investigated in the range from $0.1$ to $1$.
In addition, the inflation factor $\alpha^{rtpp}$ of RTPP is also examined in the range from $0.1$ to $1$.

In R-localization, the following Gaussian function is used as the localization function:
\begin{equation}
    L(d)=\left\{
    \begin{array}{cc}
        \exp \left(-\frac{d^{2}}{2 \sigma^{2}}\right) & \text{if} \ d<2 \sigma \sqrt{\frac{10}{3}} \\
        0 & \text{otherwise}
    \end{array}\right.,
\end{equation}
where $d$ denotes the distance between an analysis grid point and an observation, and $\sigma$ is the localization scale. 
In the URDA experiment, $\sigma$ ranges from $1$ to $10$.
When advective localization is applied, the shift speed of the localization center is set to $-0.6$ grid points per day, based on \citet{kalnay_2012}. 
Note, however, that the sign of the shift direction is reversed because URDA applies R-localization centered on grid points, unlike EFSO.
The time slots are equally divided at intervals corresponding to $1$ day.

\subsection{Evaluation metrics}
Each approach of URDA is statistically evaluated using multiple evaluation metrics. Root mean square error (RMSE) and spread are used as general evaluation metrics to assess forecast accuracy and the uncertainty of ensemble forecasts. 
Furthermore, this study introduces an evaluation metric referred to as the lead-time advantage (LTA) to quantify how long the improvement of the URDA forecast at each forecast reference time $j$ over the baseline forecast is sustained.

To define LTA, we first define the improvement rate $I_{k\mid j}$ of the URDA forecast $\mathbf{X}^{f}_{k\mid j}$ relative to the baseline forecast $\mathbf{X}^{f}_{k\mid 0}$ as follows:
\begin{equation}
    I_{k\mid j}\coloneqq\frac{\text{RMSE}^{base}_{k\mid j}-\text{RMSE}^{urda}_{k\mid j}}{\text{RMSE}^{base}_{k\mid j}}\times100\quad\left[\%\right],
\end{equation}
where $\text{RMSE}^{base}_{k\mid j}$ is the RMSE of the baseline forecast $\mathbf{X}^{f}_{k\mid 0}$, and $\text{RMSE}^{urda}_{k\mid j}$ is the RMSE of the URDA forecast $\mathbf{X}^{f}_{k\mid j}$. 
The LTA for an improvement rate $r$ is defined as the longest $k-j$ satisfying $I_{k\mid j}\geq r$:
\begin{equation}
    \text{LTA}\left(j,r\right)\coloneqq\text{max}\left\{ k-j\mid I_{k\mid j}\geq r\right\}.
\end{equation}
Note that LTA is undefined when $\left\{ k-j\mid I_{k\mid j}\geq r\right\}$ is empty. 
Although the end of the time interval $T$ is set to $7$ days in this study, the LTA for a low improvement rate can exceed $7$ days.
Therefore, we used a sufficiently long forecast lead time (14 days) only for the evaluation of LTA.
    \section{RESULTS}\label{sec:result}
\subsection{An example of URDA}\label{sec:an_example_urda}
Here, we present the results without inflation and standard R-localization ($\sigma=9.0$) as a representative example of URDA in the nonlinear model.

\begin{figure}[H]
    \centering
    \includegraphics[width=13cm]{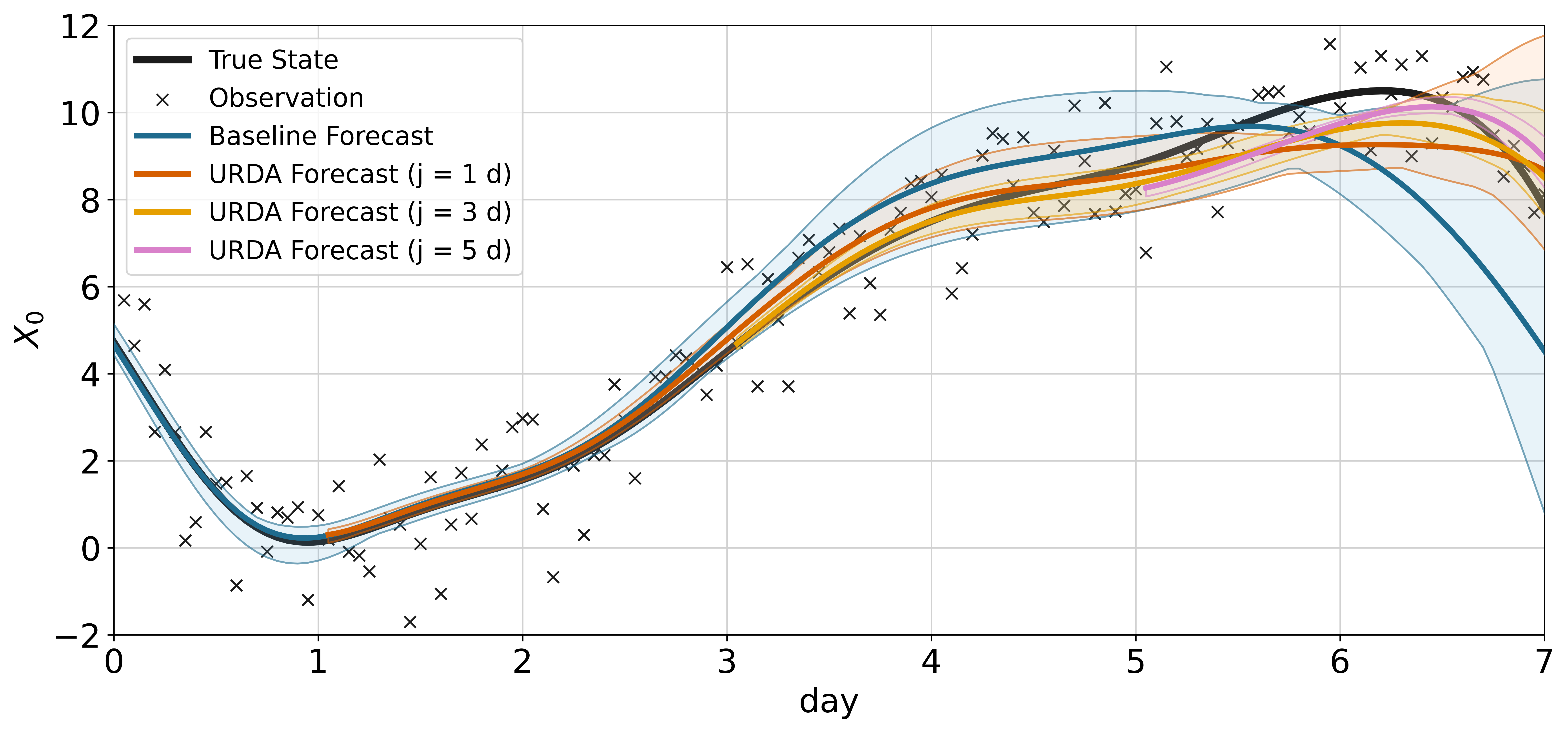}
    \caption{An example of the true state, observations, baseline forecast, and URDA forecasts without inflation and fixed localization ($\sigma=9.0$) for $g=0$ at an arbitrary initial time. The black solid line indicates the true state, black crosses indicate observations, the blue solid line indicates the baseline forecast, and the orange, yellow, and pink solid lines indicate the URDA forecasts for forecast reference times $j=1, 3$, and $5$ days, respectively. The shaded regions of each color indicate the upper and lower limits of the corresponding ensemble forecast.}
    \label{fig:urda-noinf_example}
\end{figure}

Figure \ref{fig:urda-noinf_example} illustrates an example of the URDA forecasts for $g=0$ under an arbitrary initial time, together with the true state, observations, and baseline forecast.
The baseline forecast remains close to the true state until around day 2.5, but its deviation gradually increases thereafter, resulting in a pronounced error near day 5.5.
In contrast, the URDA forecasts is generally closer to the true state than the baseline forecast, indicating an overall improvement in forecast accuracy.
However, the spread of the URDA forecasts---more precisely the range between the upper and lower bounds of the ensemble members---progressively collapses as updates are performed at each forecast reference time. In particular, at $j=5$ days, the spread is almost absent, undermining the utility as an ensemble forecast.

\begin{figure}[H]
    \centering
    \includegraphics[width=14cm]{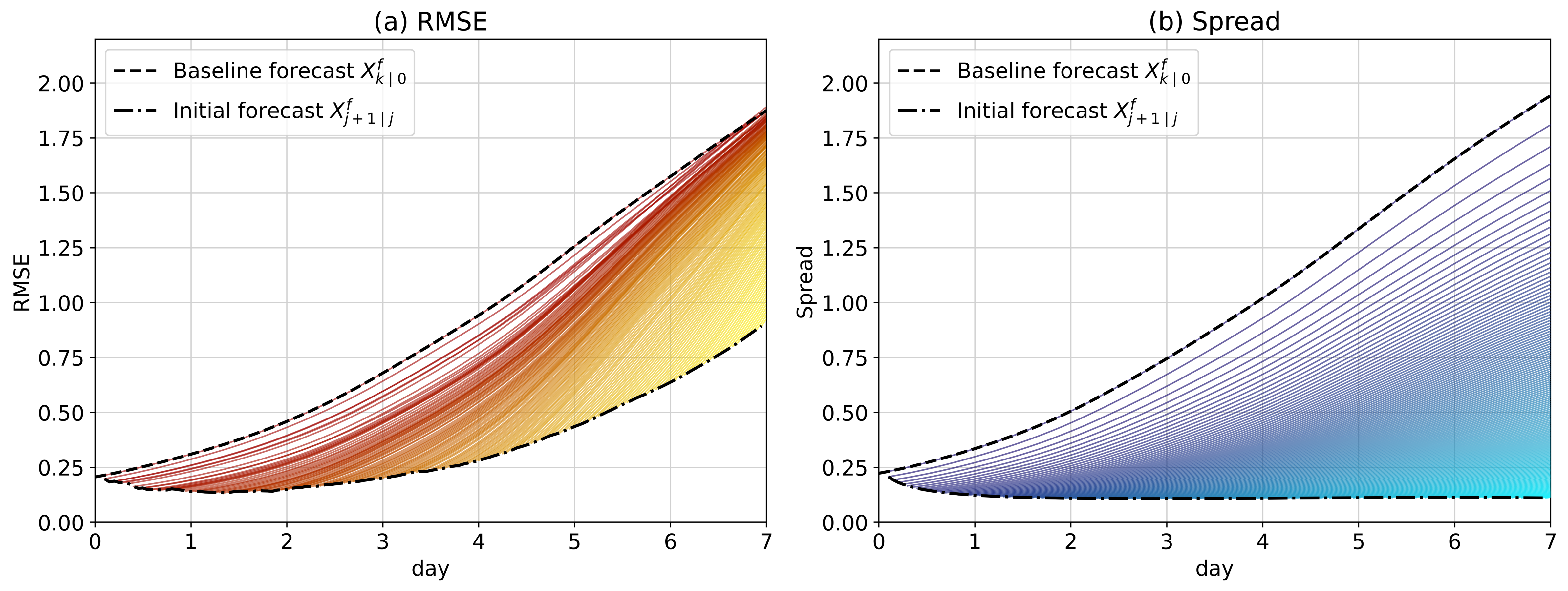}
    \caption{The RMSE and the spread of the URDA forecast without inflation and fixed localization ($\sigma=9.0$). (a) Dark and light red solid lines show the RMSE of the URDA forecast $\mathbf{X}_{k \mid j}^{f}$ for the earlier and later parts of the forecast reference time, respectively. (b) As in (a), but for the spread, shown by dark and light blue solid lines. In both panels, the dashed line represents the baseline forecast $\mathbf{X}_{k \mid 0}^{f}$, and the dash-dotted line represents the initial forecast $\mathbf{X}_{j+1 \mid j}^{f}$ for each forecast reference time.}
    \label{fig:urda-noinf_rmse_spread}
\end{figure}

Although Figure \ref{fig:urda-noinf_example} shows the results of a single example case, these tendencies can also be confirmed in the statistical evaluation shown in Figure \ref{fig:urda-noinf_rmse_spread}. 
Figure \ref{fig:urda-noinf_rmse_spread} shows the RMSE and spread of the URDA forecast.
Overall, the RMSE of the URDA forecast is improved compared to the baseline forecast. 
However, each time the URDA forecast is updated, the forecast accuracy deteriorates at a faster rate with respect to the forecast lead time, and the spread collapses.

\begin{figure}[H]
    \centering
    \includegraphics[width=8cm]{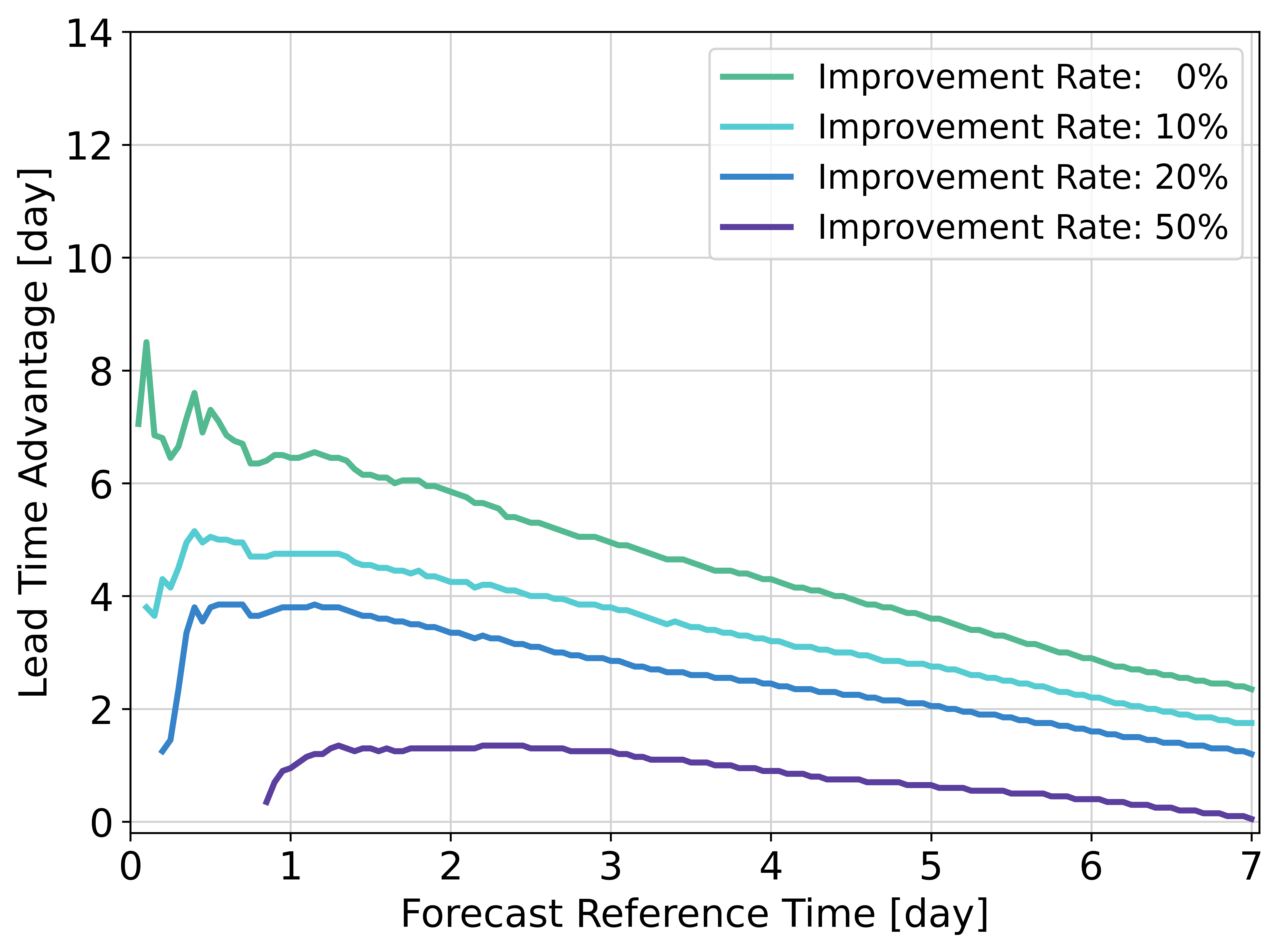}
    \caption{The LTA of the URDA forecast without inflation and fixed localization ($\sigma=9.0$). The solid green, light blue, blue, and purple lines indicate the LTA for improvement rates of $r=0, 10, 20$, and $50\%$, respectively.}
    \label{fig:urda-noinf_lta}
\end{figure}

Furthermore, we quantify the improvement of the URDA forecast over the baseline forecast using the LTA.
The LTA of the URDA forecast for improvement rates of $r=0, 10, 20$, and $50\%$ is shown in Figure \ref{fig:urda-noinf_lta}.
The LTA for an improvement rate of $r=0\%$ represents the range of $k-j$ over which the URDA forecast is no worse than the baseline forecast.
In general, the LTA tends to be longer at earlier forecast reference times and to decrease toward later forecast reference times.
However, for $r \geq 10\%$, the LTA is undefined at earlier forecast reference times because the improvement of the URDA forecast over the baseline is small, as shown in Figure \ref{fig:urda-noinf_rmse_spread}a; after the LTA becomes defined, it initially increases and then decreases.

\subsection{Inflation for URDA}\label{sec:urda-inf}
Subsequently, we investigate the properties of multiplicative inflation and RTPP in URDA. 
As representative cases, the inflation factor is set to $\alpha^{multi}=0.2$ for multiplicative inflation and $\alpha^{rtpp}=0.94$ for RTPP.

\begin{figure}[H]
    \centering
    \includegraphics[width=12cm]{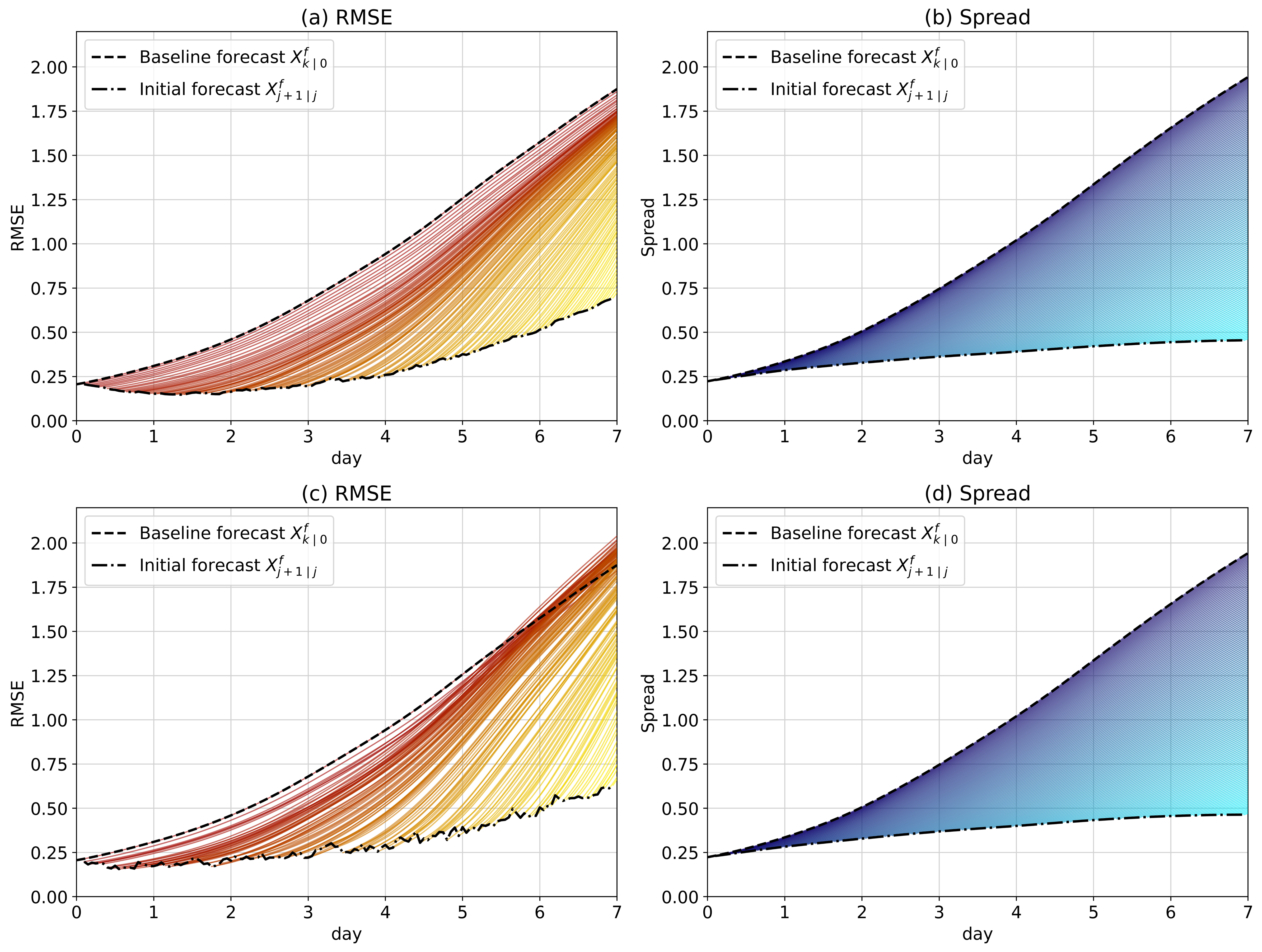}
    \caption{As in Figure \ref{fig:urda-noinf_rmse_spread}, but for (a, b) multiplicative inflation ($\alpha^{multi}=0.2$); (c, d) RTPP ($\alpha^{rtpp}=0.94$).}
    \label{fig:urda-inf_rmse_spread}
\end{figure}

The RMSE and spread of the URDA forecast with multiplicative inflation are shown in Figure \ref{fig:urda-inf_rmse_spread}a, b, while those of the URDA forecast with RTPP are shown in Figure \ref{fig:urda-inf_rmse_spread}c, d.
Regarding RMSE, the URDA forecast with multiplicative inflation exhibits an overall improvement over the baseline forecast, similar to that shown in Figure \ref{fig:urda-noinf_rmse_spread}a.
Furthermore, compared with that in Figure \ref{fig:urda-noinf_rmse_spread}a, it seems that the degradation of forecast accuracy in the URDA forecast with multiplicative inflation proceeds more gradually.
In addition, although $\alpha^{multi}=0.2$ induces deflation for $\mathbf{Y}^{f,loc}_{j\mid j-1}$, the spread is moderately inflated compared with that in Figure \ref{fig:urda-noinf_rmse_spread}b.
In contrast, although the URDA forecast with RTPP shows a spread expansion almost comparable to that of the URDA forecast with multiplicative inflation, its RMSE deteriorates more rapidly.

\begin{figure}[H]
    \centering
    \includegraphics[width=12cm]{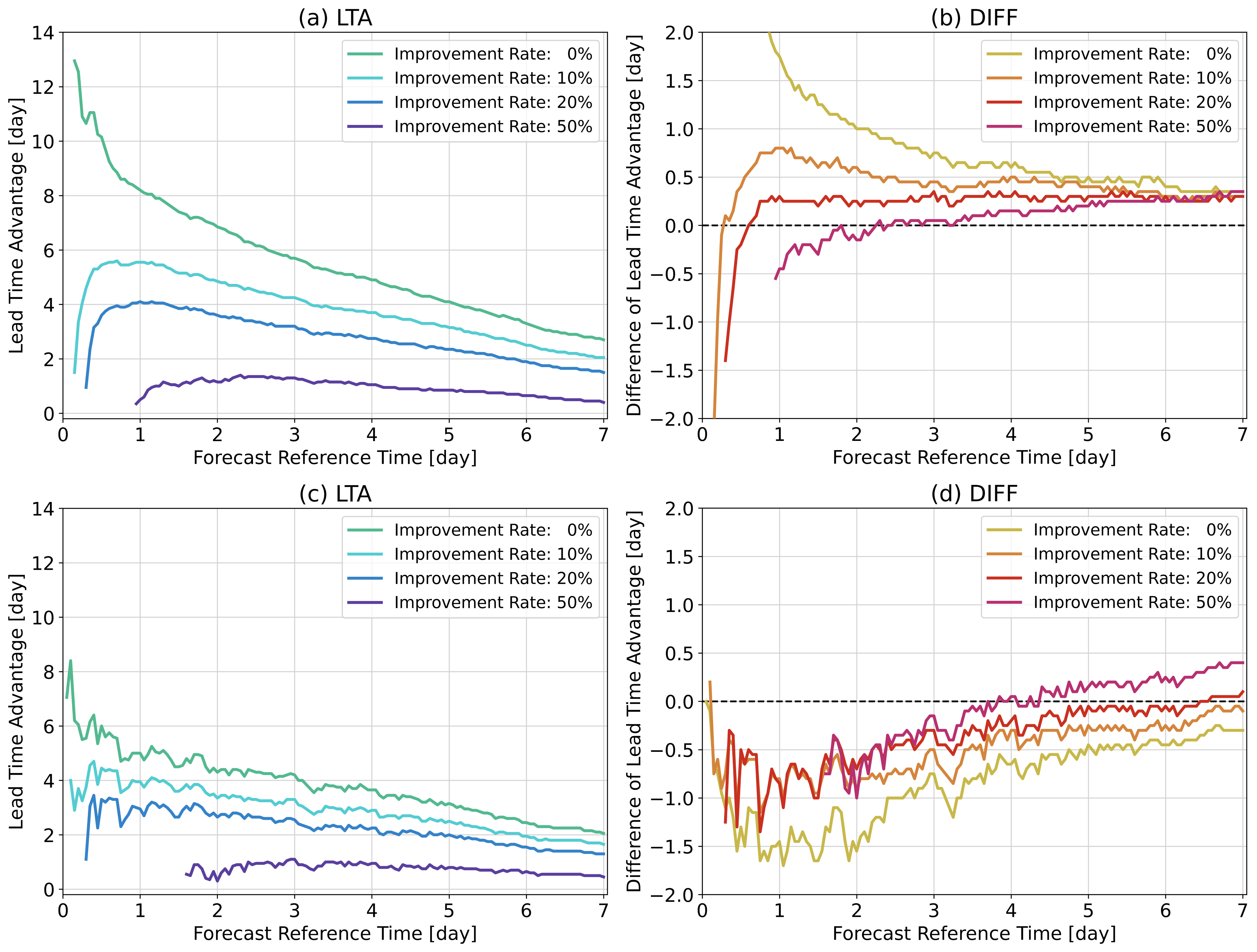}
    \caption{(a, c) As in Figure \ref{fig:urda-noinf_lta}, but for multiplicative inflation ($\alpha^{multi}=0.2$) and RTPP ($\alpha^{rtpp}=0.94$), respectively.
    (b, d) LTA differences of (a) and (c) minus no inflation, respectively; yellow, orange, red, and pink solid lines denote improvement rates of $r=0, 10, 20$, and $50\%$.}
    \label{fig:urda-inf_lta}
\end{figure}

As in the case without inflation, the improvement of the URDA forecast over the baseline forecast is quantitatively evaluated using the LTA.
Figure \ref{fig:urda-inf_lta}a,b shows the LTA of the URDA forecasts with multiplicative inflation and the difference in LTA from the no-inflation case (Figure \ref{fig:urda-noinf_lta}). The corresponding results for RTPP are shown in Figure \ref{fig:urda-inf_lta}c,d.
Applying multiplicative inflation generally improves the LTA compared to the no-inflation case.
However, some degradation is observed until a forecast reference time of approximately $2$ days.
As shown in Figure \ref{fig:urda-noinf_rmse_spread}a and Figure \ref{fig:urda-inf_rmse_spread}a, this indicates that the accuracy improvement in the URDA forecasts with multiplicative inflation is delayed at the corresponding forecast reference times.
In contrast, applying RTPP leads to an overall degradation, although slight improvements are observed for improvement rates of $20$ and $50\%$ after a forecast reference time of approximately $4$ days.

\begin{figure}[H]
    \centering
    \includegraphics[width=13cm]{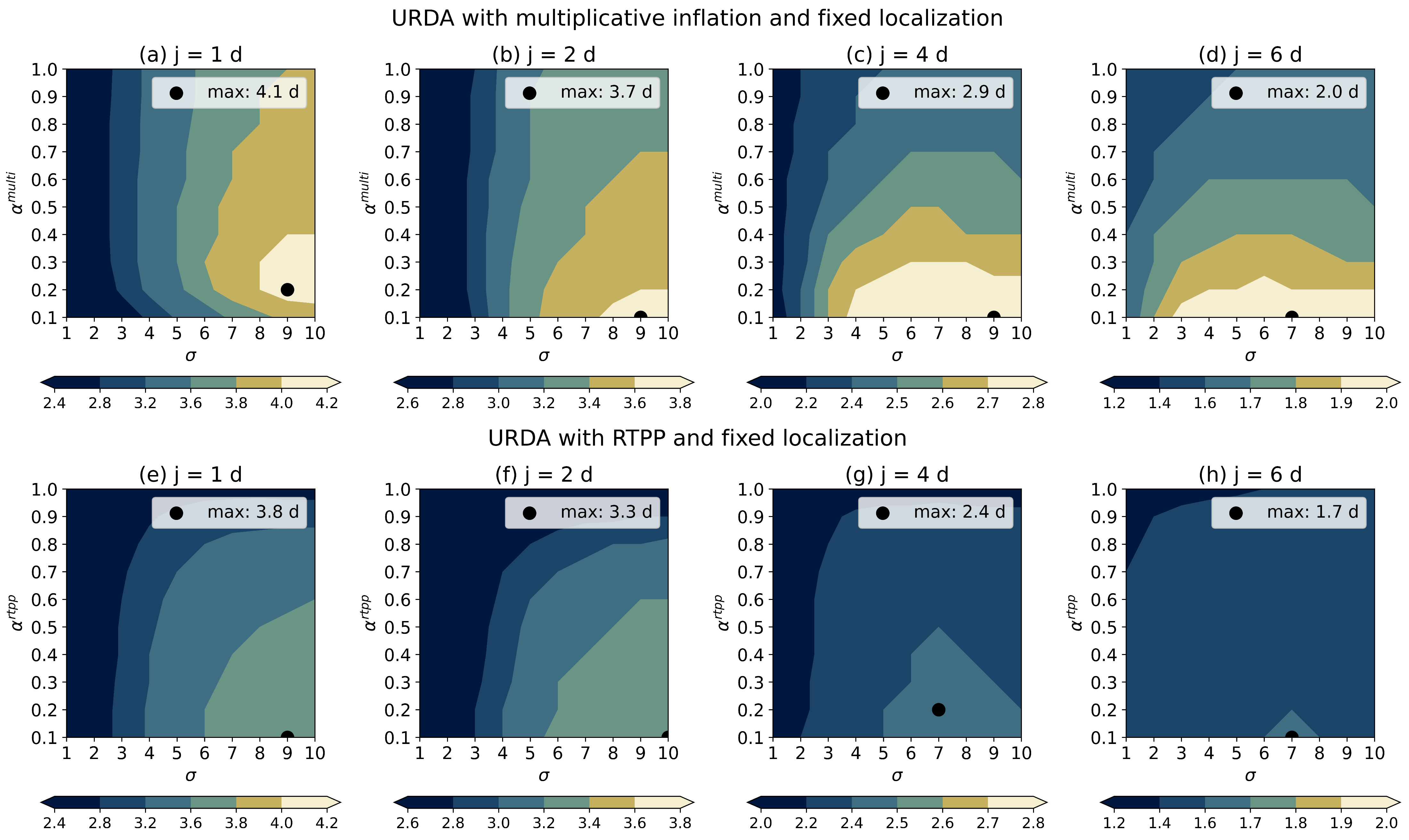}
    \caption{(a--d) The LTA for the improvement rate of $r=20\%$ for the forecast reference times of $j=1, 2, 4, 6$ days with respect to the multiplicative inflation factor $\alpha^{multi}$ and the localization scale $\sigma$. The black circle indicates the parameters that yield the best LTA. (e--h) As in (a--d), but for the inflation factor of RTPP $\alpha^{rtpp}$.}
    \label{fig:urda-inf_sensitivity}
\end{figure}

Furthermore, to investigate the tendencies of the LTA with respect to various parameter values, sensitivity experiments of the LTA are conducted for the multiplicative inflation factor $\alpha^{multi}$, the inflation factor for RTPP $\alpha^{rtpp}$, and the localization scale $\sigma$.
Figure \ref{fig:urda-inf_sensitivity}a--d shows the LTA for an improvement rate of $r=20\%$ with respect to the multiplicative inflation factor $\alpha^{multi}$ and the localization scale $\sigma$, whereas Figure \ref{fig:urda-inf_sensitivity}e--h shows the corresponding results with respect to the inflation factor for RTPP $\alpha^{rtpp}$ and the localization scale $\sigma$.
In general, multiplicative inflation yields better LTA than RTPP.
For the inflation factor, both multiplicative inflation and RTPP achieve the best LTA at values of $0.1$ or $0.2$.
This implies that, for multiplicative inflation, further deflation of $\mathbf{Y}^{f,loc}_{j\mid j-1}$ is beneficial, whereas for RTPP, it is preferable not to relax the perturbation toward those of the baseline forecast.
In addition, with respect to the localization scale $\sigma$, the best LTA is obtained for relatively large values ranging from $7$ to $10$ at each forecast reference time.

\begin{figure}[H]
    \centering
    \includegraphics[width=12cm]{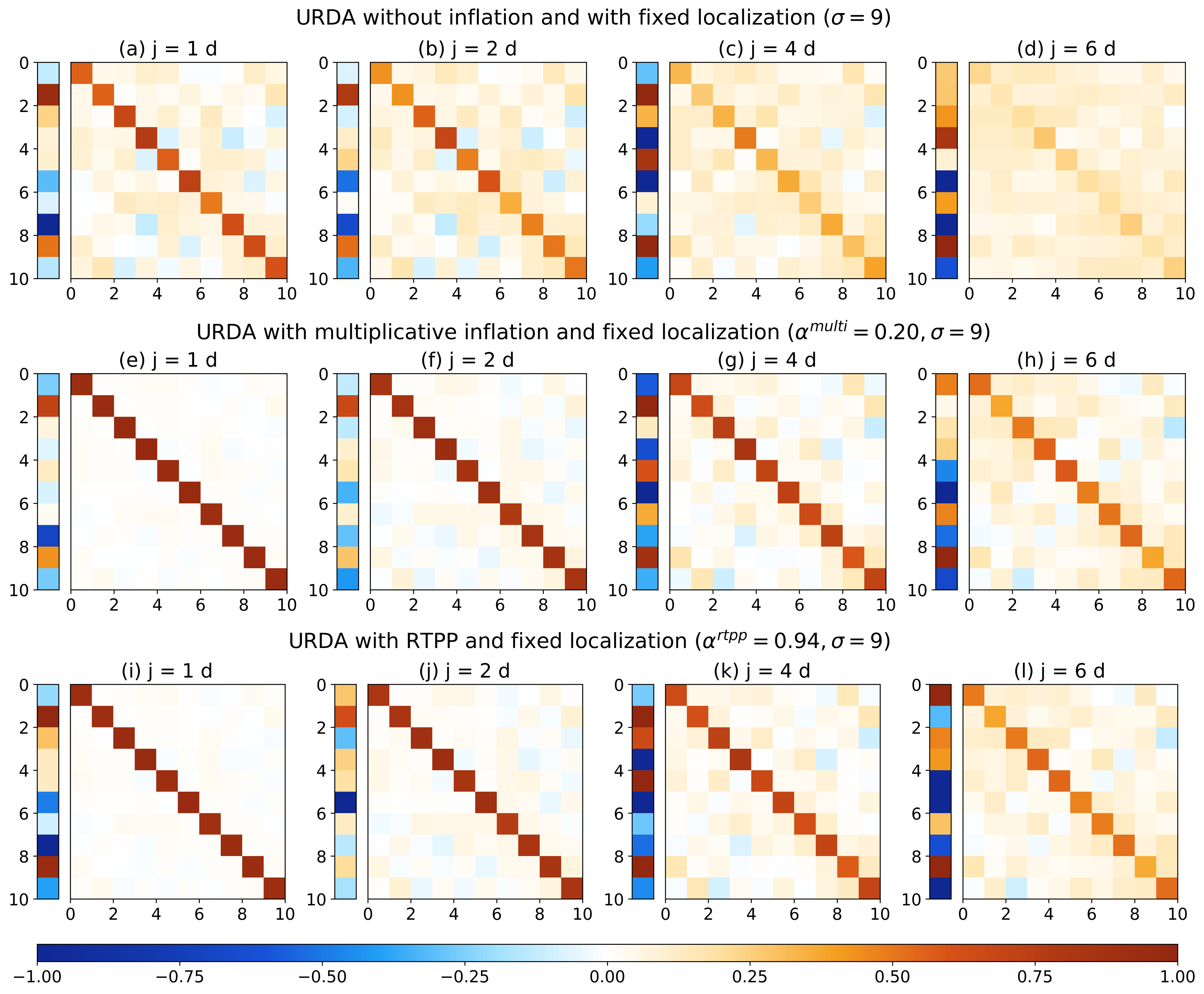}
    \caption{(a--d) The URDA forecast without inflation and with fixed localization, showing $\mathbf{w}^{prod,loc}_{j}$ and $\mathbf{W}^{prod,loc}_{j}$ at grid point $g=0$ for forecast reference times of $j=1, 2, 4, 6$ days. (e--h) As in (a--d), but for $\mathbf{w}^{prod,multi,loc}_{j}$ and $\mathbf{W}^{prod,multi,loc}_{j}$ with multiplicative inflation ($\alpha^{multi}=0.2$). (i--l) As in (a--d), but for $\mathbf{w}^{prod,rtpp,loc}_{j}$ and $\mathbf{W}^{prod,rtpp,loc}_{j}$ with RTPP ($\alpha^{rtpp}=0.94$).}
    \label{fig:urda-inf_etm}
\end{figure}

The URDA forecast is fundamentally determined by the ensemble transform matrix $\widecheck{\mathbf{W}}_{j}^{prod,loc}$, as shown in Equation \eqref{eq:urda_nonlinear_localization}. 
To understand the differences in these results arising from the inflation techniques, we decompose the ensemble transform matrix $\widecheck{\mathbf{W}}_{j}^{prod,loc}$ into $\mathbf{w}^{prod,loc}_{j}$ and $\mathbf{W}^{prod,loc}_{j}$. 
As an example, for $g=0$, $\mathbf{w}^{prod,loc}_{j}$ and $\mathbf{W}^{prod,loc}_{j}$ for no inflation, $\mathbf{w}^{prod,multi,loc}_{j}$ and $\mathbf{W}^{prod,multi,loc}_{j}$ for multiplicative inflation ($\alpha^{multi}=0.2$), and $\mathbf{w}^{prod,rtpp,loc}_{j}$ and $\mathbf{W}^{prod,rtpp,loc}_{j}$ for RTPP ($\alpha^{rtpp}=0.94$) are shown in Figure \ref{fig:urda-inf_etm}a--d, e--h, and i--l, respectively. 
Regarding $\mathbf{W}^{prod,loc}_{j}$ without inflation, moderate diagonal predominance is observed at forecast reference time $j=1$ day. However, this predominance of the diagonal elements is gradually lost as the forecast reference time progresses, and at $j=6$ days, each row becomes nearly uniform. 
In contrast, $\mathbf{W}^{prod,multi,loc}_{j}$ with multiplicative inflation ($\alpha^{multi}=0.2$) largely retains the relative predominance of the diagonal elements even as the forecast reference time progresses. 
In addition, $\mathbf{W}^{prod,rtpp,loc}_{j}$ with RTPP ($\alpha^{rtpp}=0.94$) is almost identical to $\mathbf{W}^{prod,multi,loc}_{j}$ with $\alpha^{multi}=0.2$. 
With respect to $\mathbf{w}^{prod,loc}_{j}$, $\mathbf{w}^{prod,multi,loc}_{j}$, and $\mathbf{w}^{prod,rtpp,loc}_{j}$, the results differ among the inflation techniques.

\subsection{Localization for URDA}\label{sec:urda-loc}
\begin{figure}[H]
    \centering
    \includegraphics[width=14cm]{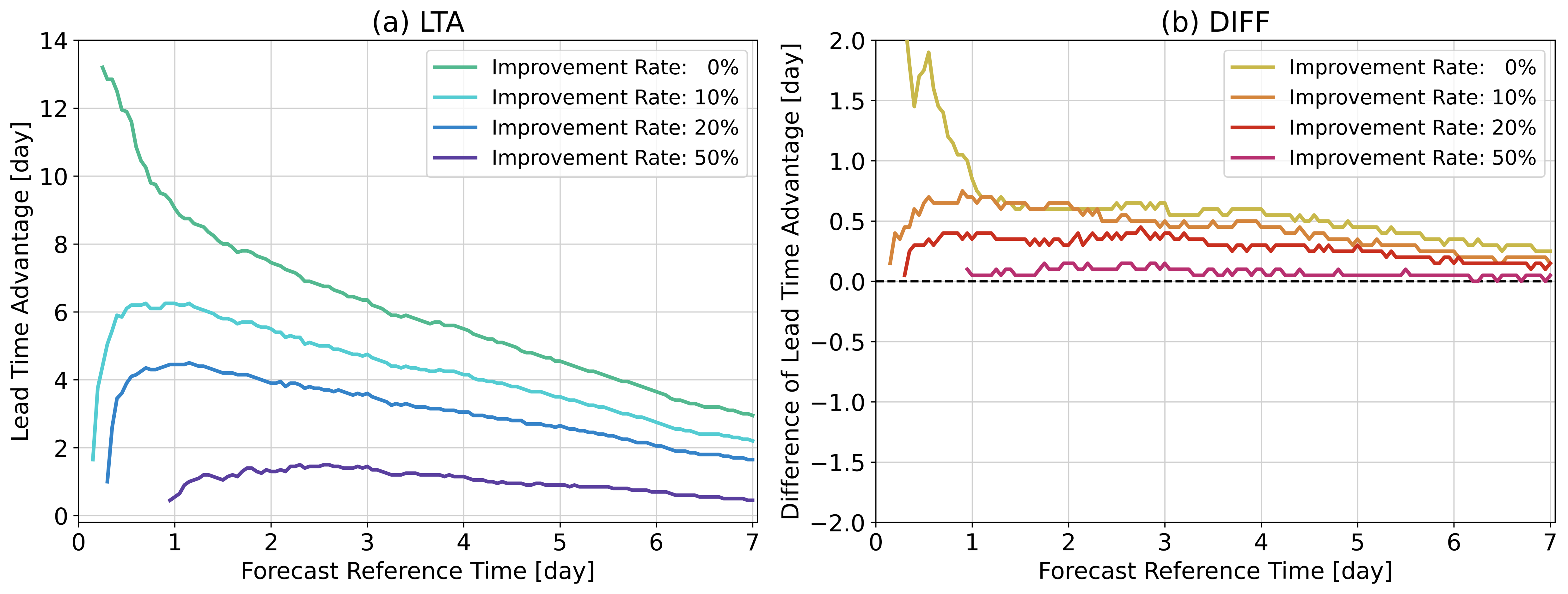}
    \caption{(a) As in Figure \ref{fig:urda-noinf_lta}, but for the URDA forecast with multiplicative inflation ($\alpha^{multi}=0.2$) and advective localization ($\sigma=9.0$). (b) As in Figure \ref{fig:urda-inf_lta}b, but for LTA differences of advective localization ($\sigma=9.0$) minus standard R-localization ($\sigma=9.0$).}
    \label{fig:urda-loc_lta}
\end{figure}

\begin{figure}[H]
    \centering
    \includegraphics[width=14cm]{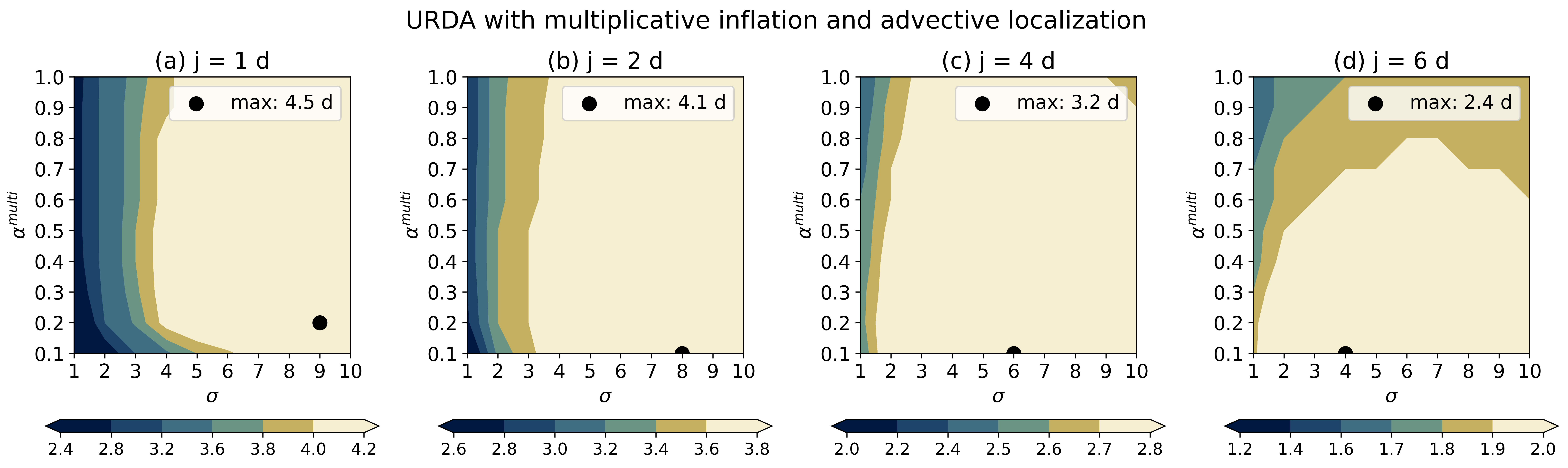}
    \caption{As in Figure \ref{fig:urda-inf_sensitivity}a--d, but for advective localization ($\sigma=9.0$).}
    \label{fig:urda-loc_sensitivity}
\end{figure}

Finally, we demonstrate the effect of advective localization in URDA.
Figure \ref{fig:urda-loc_lta} shows the LTA of the URDA forecast with multiplicative inflation ($\alpha^{multi}=0.2$) and advective localization ($\sigma=9.0$), together with the difference in LTA from the case using multiplicative inflation ($\alpha^{multi}=0.2$) and standard R-localization ($\sigma=9.0$).
The application of advective localization improves the LTA compared to standard R-localization for all forecast reference times and all improvement rates.
Furthermore, Figure \ref{fig:urda-loc_sensitivity} shows the LTA for an improvement rate of $r=20\%$ with respect to the multiplicative inflation factor $\alpha^{multi}$ and the localization scale $\sigma$ of advective localization.
Compared to the standard R-localization (Figure \ref{fig:urda-inf_sensitivity}a--d), advective localization leads to an overall improvement in the LTA.
The best LTA at each forecast reference time shows an improvement of approximately $0.3$ to $0.4$ days. 
The tendency with respect to the multiplicative inflation factor $\alpha^{multi}$ is similar to that of standard R-localization, whereas the best LTA is obtained at smaller values of the localization scale $\sigma$, ranging from $4$ to $9$ at each forecast reference time.

    \section{DISCUSSION}\label{sec:discussion}
\subsection{Effect of inflation}
Figure \ref{fig:urda-noinf_rmse_spread} shows that, without inflation, the spread collapses significantly as the forecast reference time progresses. 
The collapse of the spread implies that each row of the ensemble transform matrix $\mathbf{W}^{prod,loc}_{j}$ becomes more uniform; that is, $\mathbf{W}^{prod,loc}_{j}$ deviates further from the identity matrix.
Based on the results illustrated in Figure \ref{fig:urda-inf_etm}a--d, such uniformization of each row of the ensemble transform matrix is presumably caused by the successive multiplication of ensemble transform matrices, which is a process specific to URDA.
\citet{duc_2020} argues that the ensemble transform matrix $\mathbf{W}^{loc}_{j}$ approaches the identity matrix when the influence of observations is small, whereas the relative predominance of the diagonal elements is lost when the influence of observations is large. 
Therefore, the successive multiplication of ensemble transform matrices in URDA may reflect a process in which the influence of observations progressively accumulates.
Furthermore, we attempt to provide an interpretation from the perspective of the degrees of freedom for signal (DFS), which is a measure of the amount of information extracted from observations through data assimilation.
According to \citet{hotta_2021b}, in ETKF, excessive use of observational information---more precisely, when the true DFS obtained under ideal conditions exceeds the ensemble size---induces a collapse of the spread in the assimilated ensemble.
This theory may also be applicable to URDA.
If the successive multiplication of ensemble transform matrices in URDA indeed reflects the accumulation of the influence of observations, it can also be interpreted as a cumulative use of observational information, and the excessive use of such information could be regarded as one possible factor inducing the significant collapse of the spread accompanying the update of URDA forecasts.
However, this consideration is largely speculative reasoning, and its verification remains a subject for future investigation.

On the other hand, Figure \ref{fig:urda-inf_rmse_spread}a, b demonstrates that using $\alpha^{multi}=0.2$, which is smaller than $1$ (i.e., deflating $\mathbf{Y}^{f,loc}_{j\mid j-1}$), results in a moderate inflation of the spread.
This is because, according to Equations \eqref{eq:ensemble_transform_matrix_inf} and \eqref{eq:ensemble_analysis_error_covariance_matrix_inf}, the ensemble transform matrix $\mathbf{W}^{multi}_{j}$ approaches the identity matrix when $\alpha^{multi}<1$.
That is, the deflation of $\mathbf{Y}^{f,loc}_{j\mid j-1}$ should have an effect similar to that of RTPP.
Consequently, as shown in Figure \ref{fig:urda-inf_etm}e--h, $\mathbf{W}^{prod,multi,loc}_{j}$ is maintained moderately close to the identity matrix, and the relative predominance of its diagonal elements is preserved, thereby suppressing the collapse of the spread.
Conversely, when $\alpha^{multi}>1$, $\mathbf{W}^{multi}_{j}$ deviates further from the identity matrix, and each row of $\mathbf{W}^{prod,multi,loc}_{j}$ is expected to become uniform more rapidly.
Regarding RTPP, as shown in Figure \ref{fig:urda-inf_etm}i--l, $\mathbf{W}^{prod,rtpp,loc}_{j}$ with $\alpha^{rtpp}=0.94$ almost reproduces $\mathbf{W}^{prod,multi,loc}_{j}$ with $\alpha^{multi}=0.2$.
This again indicates that the deflation of $\mathbf{Y}^{f,loc}_{j\mid j-1}$ and RTPP have similar effects.
As a result, as shown in Figure \ref{fig:urda-inf_rmse_spread}b, d, the spread exhibits nearly identical behavior.

However, regarding forecast accuracy, particularly for the LTA, the deflation of $\mathbf{Y}^{f,loc}_{j\mid j-1}$ outperforms RTPP, as shown in Figure \ref{fig:urda-inf_sensitivity}.
We speculate that the deflation of $\mathbf{Y}^{f,loc}_{j\mid j-1}$ suppresses the errors of the tangent linear approximation in nonlinear URDA, and that this suppression contributes to the improvement in the LTA.
As shown in Lemma \ref{lem:forecast_ensemble} in Section \ref{sec:urda}, the errors of the tangent linear approximation around $\mathbf{x}^{f(i)}_{j\mid j-1}$ depend on $\sum^{m}_{l=1}\delta\mathbf{x}^{f(l)}_{j\mid j-1}\omega_{j,li}$.
To suppress these errors, $\boldsymbol{\Omega}^{loc}_{j}=\widecheck{\mathbf{W}}^{loc}_{j}-\mathbf{I}$ needs to be brought close to the zero matrix. 
That is, reducing the norm of $\mathbf{w}^{loc}_{j}$ and bringing $\mathbf{W}^{loc}_{j}$ closer to the identity matrix lead to the suppression of the errors of the tangent linear approximation in URDA forecasts.
As described above, both the deflation of $\mathbf{Y}^{f,loc}_{j\mid j-1}$ and RTPP have the effect of bringing $\mathbf{W}^{loc}_{j}$ closer to the identity matrix, and $\mathbf{W}^{multi, loc}_{j}$ with $\alpha^{multi}=0.2$ and $\mathbf{W}^{rtpp, loc}_{j}$ with $\alpha^{rtpp}=0.94$ can be nearly identical.
Therefore, the difference between the deflation and RTPP should be attributed to $\mathbf{w}^{multi,loc}_{j}$ and $\mathbf{w}^{rtpp,loc}_{j}$.
This difference arises because RTPP affects only the ensemble perturbations, whereas the deflation acts not only on the ensemble perturbations but also on the increment of the ensemble mean as given by Equations \eqref{eq:ensemble_analysis_increment_inf} and \eqref{eq:ensemble_analysis_error_covariance_matrix_inf}.
Specifically, the deflation of $\mathbf{Y}^{f,loc}_{j\mid j-1}$ directly reduces the norm of $\mathbf{w}^{multi, loc}_{j}$.
Therefore, the deflation of $\mathbf{Y}^{f,loc}_{j\mid j-1}$ can be interpreted as being effective in suppressing the errors of the tangent linear approximation in URDA forecasts, thereby leading to the improvement in the LTA.
Furthermore, as shown in Figure \ref{fig:urda-inf_lta}d, RTPP generally yields a worse LTA than no inflation, which may suggest that $\mathbf{w}^{loc}_{j}$ plays a more dominant role than $\mathbf{W}^{loc}_{j}$ in determining the errors of the tangent linear approximation, at least in the URDA experiments.
In the case of no inflation, similarly to the deflation, the norm of $\mathbf{w}^{loc}_{j}$ also decreases as the spread collapses; hence, the errors of the tangent linear approximation may be partially alleviated.
Nevertheless, as shown in Figure \ref{fig:urda-inf_sensitivity}a-d, the generally inferior LTA compared to the deflation of $\mathbf{Y}^{f,loc}_{j\mid j-1}$ is likely attributable to the fact that $\mathbf{W}^{loc}_{j}$ deviates further from the identity matrix. 
Moreover, in the case of no inflation, the collapse of the spread has not yet been significant at earlier forecast reference times, as shown in Figure \ref{fig:urda-noinf_rmse_spread}b.
Accordingly, at earlier forecast reference times, the norm of $\mathbf{w}^{loc}_{j}$ without inflation, and consequently the errors of the tangent linear approximation, may be relatively larger than that under the direct deflation of $\mathbf{Y}^{f,loc}_{j\mid j-1}$.
Since these errors accumulate over subsequent forecast reference times, aggressively deflating $\mathbf{Y}^{f,loc}_{j\mid j-1}$ from the outset may ultimately yield a better LTA.

Looking ahead to applications in NWP, when URDA is applied to the atmosphere, which has high nonlinearity, the approach of suppressing the errors of the tangent linear approximation through the deflation of $\mathbf{Y}^{f,loc}_{j\mid j-1}$ would be an effective strategy.
However, one limitation of this multiplicative inflation is that the increase in the spread diverges from the growth in the RMSE.
As shown in Figure \ref{fig:urda-inf_rmse_spread}a, b, while the RMSE grows nonlinearly, the spread increases almost linearly.
Although the multiplicative inflation factor was treated as a constant in this study, allowing it to vary adaptively may potentially resolve this issue.
Furthermore, while suppressing the errors of the tangent linear approximation is expected to yield long-term and low-to-moderate improvements in forecast accuracy, this is considered to involve a trade-off against short-term and large improvements.
This can also be understood from the fact that as $\alpha^{multi}$ approaches $0$, the URDA forecast converges to the baseline forecast.
Indeed, as shown in Figure \ref{fig:urda-inf_lta}b, the improvement in the LTA with multiplicative inflation ($\alpha^{multi}=0.2$) is delayed compared with the no-inflation case, particularly for larger improvement rates.
Additionally, from Figures \ref{fig:urda-inf_sensitivity}a--d and \ref{fig:urda-loc_sensitivity}, the tendency for the effective $\alpha^{multi}$ to decrease with increasing forecast reference time should also support this argument.
Therefore, the multiplicative inflation factor should be determined appropriately in accordance with the desired timing and magnitude of the improvement.

\subsection{Effect of localization}
In URDA, localization is also crucial; however, Figure \ref{fig:urda-inf_sensitivity} shows that the effective localization scale $\sigma$ for the LTA under standard R-localization is relatively large, ranging from $7$ to $10$.
This is presumably because the influence of observations is advected as the forecast lead time progresses, such that upstream observations relative to the analysis grid point exhibit higher correlations over time.
That is, to improve the LTA, incorporating such distant observations by using a large localization scale is required.
However, even when such observations are incorporated, their impact would be underestimated through the application of the localization function due to their distance from the localization center.
In addition, a large localization scale has side effects such as increased contamination from spurious correlations and excessive collapse of the ensemble spread.
Advective localization can resolve many of these issues.
Indeed, applying advective localization leads to an overall improvement in the LTA, as shown in Figures \ref{fig:urda-loc_lta} and \ref{fig:urda-loc_sensitivity}. 
Furthermore, it is indicated that the effective localization scale for the LTA decreases to a range of $4$ to $9$.
This is presumably because shifting the localization center toward upstream would capture highly correlated observations even at a small localization scale, while eliminating contamination from spurious correlations.
In addition, Figures \ref{fig:urda-inf_sensitivity} and \ref{fig:urda-loc_sensitivity} indicate that the effective localization scale varies with the forecast reference time. 
Specifically, except for Figure \ref{fig:urda-inf_sensitivity}f, the effective localization scale tends to decrease as the forecast reference time progresses. 
This tendency would be interpreted in terms of the errors of the tangent linear approximation discussed above. 
That is, reducing the influence of assimilated observations at each forecast reference time is expected to be effective in suppressing such errors; in the context of localization, this would be achieved by decreasing the localization scale. 
Note that, unlike inflation, localization determines not only the degree of the influence of observations but also which observations are employed. Therefore, the tendency of the localization scale may be influenced by the selection effect of the observations.

Advective localization is also expected to be effective in applications of URDA to NWP.
However, it is not straightforward how advective localization should be applied in realistic NWP models, and further investigation would be required.
For example, \citet{ota_2013} employed a method in EFSO studies that shifted the localization center based on wind speed, and this approach is considered applicable.

    \section{CONCLUSION}\label{sec:conclusion}
In this study, we investigated the properties of URDA in nonlinear models and explored inflation and localization that are effective in improving the performance of URDA, toward its application to NWP.
First, we analytically demonstrated that the preemptive forecast obtained by URDA in a nonlinear model corresponds approximately to the forecast derived from the analysis under the tangent linear approximation.
We then conducted URDA experiments using the 40-variable Lorenz 96 model. 
Regarding inflation, we demonstrated that deliberately deflating the forecast ensemble perturbations used in computing the ensemble transform matrix of URDA improves forecast accuracy and moderately inflates the ensemble spread.
In particular, forecast accuracy was quantitatively evaluated in terms of the LTA, which represents the persistence of improvement of URDA forecasts over baseline forecasts.
This improvement in the LTA is presumably attributable to the fact that deflating the forecast ensemble perturbations brings the URDA ensemble transform matrix closer to the identity matrix and reduces the increment of the ensemble mean, thereby suppressing errors in the tangent linear approximation.
In contrast, although it was shown that RTPP moderately inflates the ensemble spread, it degrades forecast accuracy from the perspective of the LTA.
We also argued, albeit briefly, that RTPS is inappropriate for URDA because it violates the properties of the ensemble transform matrix of URDA.
With regard to localization, we showed that while R-localization plays an important role, advective localization is more effective for the LTA.
This is likely because advective localization accounts for the advection of the influence of observations with forecast lead time.
    \appendix
    \section{PROPERTY OF THE ENSEMBLE TRANSFORM MATRIX}\label{sec:appendix_a}
We describe some properties of the ensemble transform matrix $\widecheck{\mathbf{W}}_{j}^{prod}$. First, we show that $\widecheck{\mathbf{W}}_{j}^{prod}$ can be decomposed into $\mathbf{w}_{j}^{prod}$ and $\mathbf{W}_{j}^{prod}$.

\begin{proposition}\label{pro:decomposition_of_ensemble_transform_matrix}
    For the ensemble transform matrix $\widecheck{\mathbf{W}}_{j}^{prod}$, the following equation holds:
    \begin{equation}
        \widecheck{\mathbf{W}}_{j}^{prod}=\frac{1}{\sqrt{m-1}} \mathbf{w}_{j}^{prod} \mathbf{1}^{\top}+\mathbf{W}_{j}^{prod}. \label{eq:proposition1}
    \end{equation}
\end{proposition}

\begin{proof}
    For $j=1$, Equation \eqref{eq:proposition1} holds as follows:
    \begin{equation}
        \widecheck{\mathbf{W}}_{1}^{prod}=\widecheck{\mathbf{W}}_{1}=\frac{1}{\sqrt{m-1}} \mathbf{w}_{1} \mathbf{1}^{\top}+\mathbf{W}_{1}=\frac{1}{\sqrt{m-1}} \mathbf{w}_{1}^{prod} \mathbf{1}^{\top}+\mathbf{W}_{1}^{prod}.
    \end{equation}
    We assume that Equation \eqref{eq:proposition1} holds for $j=t$. Then, considering the case $j=t+1$, noting that $\mathbf{1}^{\top} \mathbf{w}_{j}=0$ and $\mathbf{1}^{\top} \mathbf{W}_{j}=\mathbf{1}^{\top}$, the following equation holds:
    \begin{align}
        \widecheck{\mathbf{W}}_{t+1}^{prod} & =\widecheck{\mathbf{W}}_{t}^{prod} \widecheck{\mathbf{W}}_{t+1} \notag \\
        & =\left(\frac{1}{\sqrt{m-1}} \mathbf{w}_{t}^{prod} \mathbf{1}^{\top}+\mathbf{W}_{t}^{prod}\right)\left(\frac{1}{\sqrt{m-1}} \mathbf{w}_{t+1} \mathbf{1}^{\top}+\mathbf{W}_{t+1}\right) \notag \\
        & =\frac{1}{\sqrt{m-1}}\left(\mathbf{w}_{t}^{prod}+\mathbf{W}_{t}^{prod} \mathbf{w}_{t+1}\right) \mathbf{1}^{\top}+\mathbf{W}_{t}^{prod} \mathbf{W}_{t+1} \notag \\
        & =\frac{1}{\sqrt{m-1}} \mathbf{w}_{t+1}^{prod} \mathbf{1}^{\top}+\mathbf{W}_{t+1}^{prod}.
    \end{align}
    Therefore, because it also holds for $j=t+1$, Equation \eqref{eq:proposition1} holds for all $j$.
\end{proof}
Subsequently, regarding the column sums of $\widecheck{\mathbf{W}}_{j}^{prod}$, the following proposition is presented.

\begin{proposition}
    For the ensemble transform matrix $\widecheck{\mathbf{W}}_{j}^{prod}$, the following equation holds:
    \begin{equation}
        \mathbf{1}^{\top} \widecheck{\mathbf{W}}_{j}^{prod}=\mathbf{1}^{\top}. \label{eq:proposition2}
    \end{equation}
\end{proposition}

\begin{proof}
    For the ensemble transform matrix $\widecheck{\mathbf{W}}_{j}$, noting that $\mathbf{1}^{\top} \mathbf{w}_{j}=0$ and $\mathbf{1}^{\top} \mathbf{W}_{j}=\mathbf{1}^{\top}$, the following equation holds:
    \begin{equation}
        \mathbf{1}^{\top} \widecheck{\mathbf{W}}_{j}=\frac{1}{\sqrt{m-1}} \mathbf{1}^{\top} \mathbf{w}_{j} \mathbf{1}^{\top}+\mathbf{1}^{\top} \mathbf{W}_{j}=\mathbf{1}^{\top}.
    \end{equation}
    Therefore, we have:
    \begin{equation}
        \mathbf{1}^{\top} \widecheck{\mathbf{W}}_{j}^{prod}=\mathbf{1}^{\top} \widecheck{\mathbf{W}}_{1}  \cdots \widecheck{\mathbf{W}}_{j}=\mathbf{1}^{\top} \widecheck{\mathbf{W}}_{2} \dotsm \widecheck{\mathbf{W}}_{j}=\cdots=\mathbf{1}^{\top},
    \end{equation}
    because Equation \eqref{eq:proposition2} holds.
\end{proof}
Additionally, with respect to the row sums of $\widecheck{\mathbf{W}}_{j}^{prod}$, the following proposition is derived.

\begin{proposition}\label{pro:row_sum_of_ensemble_transform_matrix}
    For the ensemble transform matrix $\widecheck{\mathbf{W}}_{j}^{prod}$, the following equation holds:
    \begin{equation}
        \widecheck{\mathbf{W}}_{j}^{prod} \mathbf{1}=\frac{m}{\sqrt{m-1}} \mathbf{w}_{t+1}^{prod}+\mathbf{1}. \label{eq:proposition3}
    \end{equation}
\end{proposition}

\begin{proof}
    From proposition \ref{pro:decomposition_of_ensemble_transform_matrix}, noting that $\mathbf{W}_{j} \mathbf{1}=\mathbf{1}$, we have:
    \begin{align}
        \widecheck{\mathbf{W}}_{j}^{prod} \mathbf{1} & =\frac{1}{\sqrt{m-1}} \mathbf{w}_{j}^{prod} \mathbf{1}^{\top} \mathbf{1}+\mathbf{W}_{j}^{prod} \mathbf{1} \notag \\
        & =\frac{m}{\sqrt{m-1}} \mathbf{w}_{j}^{prod}+\mathbf{1}.
    \end{align}
    Therefore, Equation \eqref{eq:proposition3} holds.
\end{proof}
As shown in the following corollaries, $\mathbf{w}_{j}^{prod}$ and $\mathbf{W}_{j}^{prod}$ can be expressed using $\widecheck{\mathbf{W}}_{j}^{prod}$.

\begin{corollary}\label{cor:ensemble_analysis_increment_prod}
    Using the ensemble transform matrix $\widecheck{\mathbf{W}}_{j}^{prod}$, the vector $\mathbf{w}_{j}^{prod}$ can be expressed as follows:
    \begin{equation}
        \mathbf{w}_{j}^{prod}=\frac{\sqrt{m-1}}{m}\left(\widecheck{\mathbf{W}}_{j}^{prod}-\mathbf{I}\right) \mathbf{1}.
    \end{equation}
\end{corollary}

\begin{proof}
    This follows directly from proposition \ref{pro:row_sum_of_ensemble_transform_matrix}.
\end{proof}

\begin{corollary}
    Using the ensemble transform matrix $\widecheck{\mathbf{W}}_{j}^{prod}$, the ensemble transform matrix $\mathbf{W}_{j}^{prod}$ can be expressed as follows:
    \begin{equation}
        \mathbf{W}_{j}^{prod}=\widecheck{\mathbf{W}}_{j}^{prod}-\frac{1}{m}\left(\widecheck{\mathbf{W}}_{j}^{prod}-\mathbf{I}\right) \mathbf{J}.
    \end{equation}
\end{corollary}

\begin{proof}
    This follows directly from proposition \ref{pro:decomposition_of_ensemble_transform_matrix} and corollary \ref{cor:ensemble_analysis_increment_prod}.
\end{proof}
    \section*{ACKNOWLEDGEMENTS}
The authors thank Dr. Tadashi Tsuyuki of the Meteorological Research Institute for valuable discussions and insights that contributed to this study.
In addition, the authors are deeply grateful to the editors and the anonymous reviewer of the Quarterly Journal of the Royal Meteorological Society for their invaluable comments and constructive suggestions, which significantly improved the quality of this manuscript.
In particular, the experimental design of evaluating URDA forecasts within the range where the tangent linear approximation is expected to hold, as well as the lead-time advantage (LTA) as an evaluation metric, stems from the insightful suggestions made by the anonymous reviewer.

\section*{FINANCIAL SUPPORT}
This study was supported by the Japan Society for the Promotion of Science (JSPS) through KAKENHI (Grant Numbers 25KJ0729 and 25H00752), the Japan Science and Technology Agency (JST) Moonshot Research and Development Program (Grant Number JPMJMS2389), and the Institute for Advanced Academic Research (IAAR) Research Support Program.

    \begingroup
	\normalem
	\bibliographystyle{apalike}
	\bibliography{reference.bib}

@article{anderson_1999,
  title = {A {{Monte Carlo Implementation}} of the {{Nonlinear Filtering Problem}} to {{Produce Ensemble Assimilations}} and {{Forecasts}}},
  author = {Anderson, Jeffrey L. and Anderson, Stephen L.},
  year = {1999},
  month = dec,
  journal = {Monthly Weather Review},
  volume = {127},
  number = {12},
  pages = {2741--2758},
  publisher = {American Meteorological Society},
  issn = {1520-0493, 0027-0644},
  doi = {10.1175/1520-0493(1999)127<2741:AMCIOT>2.0.CO;2},
  urldate = {2025-09-24},
  chapter = {Monthly Weather Review},
  langid = {english},
}

@article{bauer_2015,
  title = {The Quiet Revolution of Numerical Weather Prediction},
  author = {Bauer, Peter and Thorpe, Alan and Brunet, Gilbert},
  year = {2015},
  month = sep,
  journal = {Nature},
  volume = {525},
  number = {7567},
  pages = {47--55},
  publisher = {Nature Publishing Group},
  issn = {1476-4687},
  doi = {10.1038/nature14956},
  urldate = {2025-06-27},
  copyright = {2015 Springer Nature Limited},
  langid = {english},
}

@article{bishop_2001,
  title = {Adaptive {{Sampling}} with the {{Ensemble Transform Kalman Filter}}. {{Part I}}: {{Theoretical Aspects}}},
  shorttitle = {Adaptive {{Sampling}} with the {{Ensemble Transform Kalman Filter}}. {{Part I}}},
  author = {Bishop, Craig H. and Etherton, Brian J. and Majumdar, Sharanya J.},
  year = {2001},
  month = mar,
  journal = {Monthly Weather Review},
  volume = {129},
  number = {3},
  pages = {420--436},
  publisher = {American Meteorological Society},
  issn = {1520-0493, 0027-0644},
  doi = {10.1175/1520-0493(2001)129<0420:ASWTET>2.0.CO;2},
  urldate = {2025-06-27},
  chapter = {Monthly Weather Review},
  langid = {english},
}

@article{dimet_1986,
  title = {Variational Algorithms for Analysis and Assimilation of Meteorological Observations: Theoretical Aspects},
  shorttitle = {Variational Algorithms for Analysis and Assimilation of Meteorological Observations},
  author = {Dimet, Fran{\c c}ois-Xavier Le and Talagrand, Olivier},
  year = {1986},
  journal = {Tellus A},
  volume = {38A},
  number = {2},
  pages = {97--110},
  issn = {1600-0870},
  doi = {10.1111/j.1600-0870.1986.tb00459.x},
  urldate = {2025-01-13},
  copyright = {1986 Blackwell Munksgaard},
  langid = {english},
}

@article{duc_2020,
  title = {An {{Explanation}} for the {{Diagonally Predominant Property}} of the {{Positive Symmetric Ensemble Transform Matrix}}},
  author = {Duc, Le and Saito, Kazuo and Hotta, Daisuke},
  year = {2020},
  journal = {Journal of the Meteorological Society of Japan. Ser. II},
  volume = {98},
  number = {2},
  pages = {455--462},
  doi = {10.2151/jmsj.2020-022},
}

@article{etherton_2007,
  title = {Preemptive {{Forecasts Using}} an {{Ensemble Kalman Filter}}},
  author = {Etherton, Brian J.},
  year = {2007},
  month = oct,
  journal = {Monthly Weather Review},
  volume = {135},
  number = {10},
  pages = {3484--3495},
  publisher = {American Meteorological Society},
  issn = {1520-0493, 0027-0644},
  doi = {10.1175/MWR3480.1},
  urldate = {2024-09-05},
  chapter = {Monthly Weather Review},
  langid = {english},
}

@article{evensen_1994,
  title = {Sequential Data Assimilation with a Nonlinear Quasi-Geostrophic Model Using {{Monte Carlo}} Methods to Forecast Error Statistics},
  author = {Evensen, Geir},
  year = {1994},
  journal = {Journal of Geophysical Research: Oceans},
  volume = {99},
  number = {C5},
  pages = {10143--10162},
  issn = {2156-2202},
  doi = {10.1029/94JC00572},
  urldate = {2023-11-25},
  copyright = {Copyright 1994 by the American Geophysical Union.},
  langid = {english}
}

@book{evensen_2022,
  title = {Data {{Assimilation Fundamentals}}: {{A Unified Formulation}} of the {{State}} and {{Parameter Estimation Problem}}},
  shorttitle = {Data {{Assimilation Fundamentals}}},
  author = {Evensen, Geir and Vossepoel, Femke C. and Van Leeuwen, Peter Jan},
  year = {2022},
  series = {Springer {{Textbooks}} in {{Earth Sciences}}, {{Geography}} and {{Environment}}},
  publisher = {Springer International Publishing},
  address = {Cham},
  doi = {10.1007/978-3-030-96709-3},
  urldate = {2025-06-27},
  copyright = {https://creativecommons.org/licenses/by/4.0},
  isbn = {978-3-030-96708-6 978-3-030-96709-3},
  langid = {english},
}

@article{hotta_2021b,
  title = {Why Does {{EnKF}} Suffer from Analysis Overconfidence? {{An}} Insight into Exploiting the Ever-Increasing Volume of Observations},
  shorttitle = {Why Does {{EnKF}} Suffer from Analysis Overconfidence?},
  author = {Hotta, Daisuke and Ota, Yoichiro},
  year = {2021},
  journal = {Quarterly Journal of the Royal Meteorological Society},
  volume = {147},
  number = {735},
  pages = {1258--1277},
  issn = {1477-870X},
  doi = {10.1002/qj.3970},
  urldate = {2026-05-11},
  copyright = {{\copyright} 2020 Royal Meteorological Society},
  langid = {english},
}

@article{hunt_2007,
  title = {Efficient Data Assimilation for Spatiotemporal Chaos: {{A}} Local Ensemble Transform {{Kalman}} Filter},
  shorttitle = {Efficient Data Assimilation for Spatiotemporal Chaos},
  author = {Hunt, Brian R. and Kostelich, Eric J. and Szunyogh, Istvan},
  year = {2007},
  month = jun,
  journal = {Physica D: Nonlinear Phenomena},
  series = {Data {{Assimilation}}},
  volume = {230},
  number = {1},
  pages = {112--126},
  issn = {0167-2789},
  doi = {10.1016/j.physd.2006.11.008},
  urldate = {2025-06-27},
}

@book{kalnay_2002,
  title = {Atmospheric {{Modeling}}, {{Data Assimilation}} and {{Predictability}}},
  author = {Kalnay, Eugenia},
  year = {2002},
  month = nov,
  publisher = {Cambridge University Press},
  doi = {10.1017/CBO9780511802270},
  urldate = {2025-06-27},
  isbn = {9780511802270},
  langid = {english}
}

@article{kalnay_2012,
  title = {A Simpler Formulation of Forecast Sensitivity to Observations: Application to Ensemble {{Kalman}} Filters},
  shorttitle = {A Simpler Formulation of Forecast Sensitivity to Observations},
  author = {Kalnay, Eugenia and Ota, Yoichiro and Miyoshi, Takemasa and Liu, Junjie},
  year = {2012},
  month = dec,
  journal = {Tellus A: Dynamic Meteorology and Oceanography},
  volume = {64},
  number = {1},
  pages = {18462},
  publisher = {Taylor \& Francis},
  issn = {null},
  doi = {10.3402/tellusa.v64i0.18462},
  urldate = {2026-03-14},
}

@article{kotsuki_2020,
  title = {Weight Structure of the {{Local Ensemble Transform Kalman Filter}}: {{A}} Case with an Intermediate Atmospheric General Circulation Model},
  shorttitle = {Weight Structure of the {{Local Ensemble Transform Kalman Filter}}},
  author = {Kotsuki, Shunji and Pensoneault, Andrew and Okazaki, Atsushi and Miyoshi, Takemasa},
  year = {2020},
  journal = {Quarterly Journal of the Royal Meteorological Society},
  volume = {146},
  number = {732},
  pages = {3399--3415},
  issn = {1477-870X},
  doi = {10.1002/qj.3852},
  urldate = {2025-06-28},
  copyright = {{\copyright} 2020 The Authors. Quarterly Journal of the Royal Meteorological Society published by John Wiley \& Sons Ltd on behalf of the Royal Meteorological Society.},
  langid = {english},
}

@article{liu_2008,
  title = {An {{Ensemble-Based Four-Dimensional Variational Data Assimilation Scheme}}. {{Part I}}: {{Technical Formulation}} and {{Preliminary Test}}},
  shorttitle = {An {{Ensemble-Based Four-Dimensional Variational Data Assimilation Scheme}}. {{Part I}}},
  author = {Liu, Chengsi and Xiao, Qingnong and Wang, Bin},
  year = {2008},
  month = sep,
  journal = {Monthly Weather Review},
  volume = {136},
  number = {9},
  pages = {3363--3373},
  publisher = {American Meteorological Society},
  issn = {1520-0493, 0027-0644},
  doi = {10.1175/2008MWR2312.1},
  urldate = {2025-01-05},
  chapter = {Monthly Weather Review},
  langid = {english},
}

@article{lorenz_1963,
  title = {Deterministic {{Nonperiodic Flow}}},
  author = {Lorenz, Edward N.},
  year = {1963},
  month = mar,
  journal = {Journal of the Atmospheric Sciences},
  volume = {20},
  number = {2},
  pages = {130--141},
  publisher = {American Meteorological Society},
  issn = {0022-4928, 1520-0469},
  doi = {10.1175/1520-0469(1963)020<0130:DNF>2.0.CO;2},
  urldate = {2023-10-27},
  chapter = {Journal of the Atmospheric Sciences},
  langid = {english},
}

@article{lorenz_1996,
  title = {Predictability: A Problem Partly Solved},
  shorttitle = {Predictability},
  author = {Lorenz, E. N.},
  year = {1996},
  urldate = {2024-01-23},
  langid = {english},
}

@article{lorenz_1998a,
  title = {Optimal {{Sites}} for {{Supplementary Weather Observations}}: {{Simulation}} with a {{Small Model}}},
  shorttitle = {Optimal {{Sites}} for {{Supplementary Weather Observations}}},
  author = {Lorenz, Edward N. and Emanuel, Kerry A.},
  year = {1998},
  month = feb,
  journal = {Journal of the Atmospheric Sciences},
  volume = {55},
  number = {3},
  pages = {399--414},
  publisher = {American Meteorological Society},
  issn = {0022-4928, 1520-0469},
  doi = {10.1175/1520-0469(1998)055<0399:OSFSWO>2.0.CO;2},
  urldate = {2025-01-13},
  chapter = {Journal of the Atmospheric Sciences},
  langid = {english},
}

@article{madaus_2015,
  title = {Rapid, Short-Term Ensemble Forecast Adjustment through Offline Data Assimilation},
  author = {Madaus, L. E. and Hakim, G. J.},
  year = {2015},
  journal = {Quarterly Journal of the Royal Meteorological Society},
  volume = {141},
  number = {692},
  pages = {2630--2642},
  issn = {1477-870X},
  doi = {10.1002/qj.2549},
  urldate = {2024-09-05},
  copyright = {{\copyright} 2015 Royal Meteorological Society},
  langid = {english},
}

@article{miyoshi_2006,
  title = {Applying a {{Four-dimensional Local Ensemble Transform Kalman Filter}} ({{4D-LETKF}}) to the {{JMA Nonhydrostatic Model}} ({{NHM}})},
  author = {Miyoshi, Takemasa and Aranami, Kohei},
  year = {2006},
  journal = {Sola},
  volume = {2},
  pages = {128--131},
  doi = {10.2151/sola.2006-033},
}

@article{miyoshi_2007,
  title = {Local {{Ensemble Transform Kalman Filtering}} with an {{AGCM}} at a {{T159}}/{{L48 Resolution}}},
  author = {Miyoshi, Takemasa and Yamane, Shozo},
  year = {2007},
  month = nov,
  journal = {Monthly Weather Review},
  volume = {135},
  number = {11},
  pages = {3841--3861},
  publisher = {American Meteorological Society},
  issn = {1520-0493, 0027-0644},
  doi = {10.1175/2007MWR1873.1},
  urldate = {2025-06-28},
  chapter = {Monthly Weather Review},
  langid = {english},
}

@article{ota_2013,
  title = {Ensemble-Based Observation Impact Estimates Using the {{NCEP GFS}}},
  author = {Ota, Yoichiro and Derber, John C. and Kalnay, Eugenia and Miyoshi, Takemasa},
  year = {2013},
  month = dec,
  journal = {Tellus},
  volume = {65},
  number = {1},
  issn = {3035-9554},
  doi = {10.3402/tellusa.v65i0.20038},
  urldate = {2026-03-15},
  langid = {english},
}

@article{potthast_2018,
  title = {Ultra {{Rapid Data Assimilation Based}} on {{Ensemble Filters}}},
  author = {Potthast, Roland and Welzbacher, Christian A.},
  year = {2018},
  month = oct,
  journal = {Frontiers in Applied Mathematics and Statistics},
  volume = {4},
  publisher = {Frontiers},
  issn = {2297-4687},
  doi = {10.3389/fams.2018.00045},
  urldate = {2024-09-06},
  langid = {english},
}

@article{szunyogh_2008,
  title = {A Local Ensemble Transform {{Kalman}} Filter Data Assimilation System for the {{NCEP}} Global Model},
  author = {Szunyogh, Istvan and , Eric J., Kostelich and , Gyorgyi, Gyarmati and , Eugenia, Kalnay and , Brian R., Hunt and , Edward, Ott and , Elizabeth, Satterfield and {and Yorke}, James A.},
  year = {2008},
  month = jan,
  journal = {Tellus A: Dynamic Meteorology and Oceanography},
  volume = {60},
  number = {1},
  pages = {113--130},
  publisher = {Taylor \& Francis},
  issn = {null},
  doi = {10.1111/j.1600-0870.2007.00274.x},
  urldate = {2025-06-28},
}

@article{talagrand_1987,
  title = {Variational {{Assimilation}} of {{Meteorological Observations With}} the {{Adjoint Vorticity Equation}}. {{I}}: {{Theory}}},
  shorttitle = {Variational {{Assimilation}} of {{Meteorological Observations With}} the {{Adjoint Vorticity Equation}}. {{I}}},
  author = {Talagrand, Olivier and Courtier, Philippe},
  year = {1987},
  journal = {Quarterly Journal of the Royal Meteorological Society},
  volume = {113},
  number = {478},
  pages = {1311--1328},
  issn = {1477-870X},
  doi = {10.1002/qj.49711347812},
  urldate = {2025-01-13},
  copyright = {Copyright {\copyright} 1987 Royal Meteorological Society},
  langid = {english}
}

@article{whitaker_2002,
  title = {Ensemble {{Data Assimilation}} without {{Perturbed Observations}}},
  author = {Whitaker, Jeffrey S. and Hamill, Thomas M.},
  year = {2002},
  month = jul,
  journal = {Monthly Weather Review},
  volume = {130},
  number = {7},
  pages = {1913--1924},
  publisher = {American Meteorological Society},
  issn = {1520-0493, 0027-0644},
  doi = {10.1175/1520-0493(2002)130<1913:EDAWPO>2.0.CO;2},
  urldate = {2025-06-27},
  chapter = {Monthly Weather Review},
  langid = {english},
}

@article{whitaker_2012,
  title = {Evaluating {{Methods}} to {{Account}} for {{System Errors}} in {{Ensemble Data Assimilation}}},
  author = {Whitaker, Jeffrey S. and Hamill, Thomas M.},
  year = {2012},
  month = sep,
  journal = {Monthly Weather Review},
  volume = {140},
  number = {9},
  pages = {3078--3089},
  publisher = {American Meteorological Society},
  issn = {1520-0493, 0027-0644},
  doi = {10.1175/MWR-D-11-00276.1},
  urldate = {2025-09-11},
  chapter = {Monthly Weather Review},
  langid = {english},
}

@article{zhang_2004,
  title = {Impacts of {{Initial Estimate}} and {{Observation Availability}} on {{Convective-Scale Data Assimilation}} with an {{Ensemble Kalman Filter}}},
  author = {Zhang, F. and Snyder, Chris and Sun, Juanzhen},
  year = {2004},
  month = may,
  journal = {Monthly Weather Review},
  volume = {132},
  number = {5},
  pages = {1238--1253},
  publisher = {American Meteorological Society},
  issn = {1520-0493, 0027-0644},
  doi = {10.1175/1520-0493(2004)132<1238:IOIEAO>2.0.CO;2},
  urldate = {2025-09-11},
  chapter = {Monthly Weather Review},
  langid = {english},
}

@article{zhang_2023,
  title = {Skilful Nowcasting of Extreme Precipitation with {{NowcastNet}}},
  author = {Zhang, Yuchen and Long, Mingsheng and Chen, Kaiyuan and Xing, Lanxiang and Jin, Ronghua and Jordan, Michael I. and Wang, Jianmin},
  year = {2023},
  month = jul,
  journal = {Nature},
  volume = {619},
  number = {7970},
  pages = {526--532},
  publisher = {Nature Publishing Group},
  issn = {1476-4687},
  doi = {10.1038/s41586-023-06184-4},
  urldate = {2025-06-26},
  copyright = {2023 The Author(s)},
  langid = {english},
}
	\endgroup
\end{document}